\documentclass[sigconf, nonacm]{acmart}
\newcommand\vldbdoi{XX.XX/XXX.XX}
\newcommand\vldbpages{XXX-XXX}
\newcommand\vldbvolume{1}
\newcommand\vldbissue{1}
\newcommand\vldbyear{2026}
\newcommand\vldbauthors{\authors}
\newcommand\vldbtitle{\shorttitle}
\newcommand\vldbavailabilityurl{https://github.com/duerwuyi/DAT}
\newcommand\vldbpagestyle{plain}

\usepackage{enumitem}
\usepackage{tabularray}
\usepackage{multirow}
\usepackage{graphicx}
\usepackage{pifont}
\usepackage{xspace}
\usepackage{tikz}
\usepackage{algorithm}
\usepackage{algpseudocode}
\usepackage{subcaption}
\usepackage[dvipsnames]{xcolor}
\usepackage{listings} % For code snippets
\lstdefinelanguage{Cypher}
{
  morekeywords={MATCH, RETURN, WHERE, CREATE, SET, DELETE, DETACH, UNWIND, MERGE, OPTIONAL, FOREACH, UNION, WITH, AS, ORDER, BY, ASC, DESC, LIMIT, SKIP, AND, OR, NOT, XOR, TRUE, FALSE, NULL, IN, START, END, CONTAINS, ALL, ANY, NONE, SINGLE, DISTINCT, CASE, WHEN, THEN, ELSE, ENDS, EXISTS, CASE, WHEN, THEN, ELSE, END, CALL, IS},
  sensitive=true,
  morecomment=[l]{//},
  morecomment=[s]{/*}{*/},
  morestring=[b]",
  morestring=[b]',
}
\usepackage{fontawesome}  % For \faBug

\usepackage{longtable}

\usepackage{xcolor}
\usepackage{listings}

\lstdefinelanguage{jsonLanguage}{
    basicstyle=\ttfamily\footnotesize,
    numbers=left,
    numberstyle=\scriptsize,
    breaklines=true,
    frame=lines,
    backgroundcolor=\color{gray!10},
    showstringspaces=false,
    string=[db]{"},
    stringstyle=\color{green!50!black},
    morestring=[s][\color{black}]{\ \ "}{":},
    keywordstyle=\color{blue},
    keywords={true,false,null},
    morecomment=[s]{/*}{*/},
    emph={ckey},
    emphstyle=\color{orange!80!black},
    literate=
     *{0}{{{\color{red}0}}}{1}
      {1}{{{\color{red}1}}}{1}
      {2}{{{\color{red}2}}}{1}
      {3}{{{\color{red}3}}}{1}
      {4}{{{\color{red}4}}}{1}
      {5}{{{\color{red}5}}}{1}
      {6}{{{\color{red}6}}}{1}
      {7}{{{\color{red}7}}}{1}
      {8}{{{\color{red}8}}}{1}
      {9}{{{\color{red}9}}}{1}
      {"ckey"}{{\textcolor{orange!80!black}{"ckey"}}}6
      {.}{{{\color{red}.}}}{1}
      {:}{{{\color{gray}{:}}}}{1}
      {,}{{{\color{gray}{,}}}}{1}
      {\{}{{{\color{gray}{\{}}}}{1}
      {\}}{{{\color{gray}{\}}}}}{1}
      {[}{{{\color{gray}{[}}}}{1}
      {]}{{{\color{gray}{]}}}}{1},
}

\lstdefinestyle{sql}{
  language=SQL,
  basicstyle=\ttfamily\footnotesize, % 小一号
  keywordstyle=\color{blue}\bfseries,
  commentstyle=\color{black!55},
  stringstyle=\color{black},
  numbers=left,
  numberstyle=\scriptsize\color{black},
  numbersep=10pt,
  xleftmargin=1.8em,
  breaklines=true,
  columns=fullflexible,
  keepspaces=true,
  showstringspaces=false,
  tabsize=2,
  frame=none,
  emph={ckey},
  emphstyle=\color{orange!80!black},
  emph={[2]create_distributed_table,create_reference_table, SINGLE, STORAGE, UNIT, BROADCAST, RULE},
  emphstyle={[2]\color{red}\bfseries}
}

\lstdefinestyle{json}{
  language=jsonLanguage,
  basicstyle=\ttfamily\footnotesize, % 小一号
  numbers=left,
  numberstyle=\scriptsize\color{black},
  numbersep=10pt,
  xleftmargin=1.8em,
  backgroundcolor=\color{black!4},
  frame=single,
  rulecolor=\color{black!35},
  framesep=4pt,
  commentstyle=\color{black!45}, % 让 /* Vitess VSchema ... */ 变灰
  emph={ckey},
  emphstyle=\color{orange!80!black}
}

\usepackage{booktabs}     % 画表格横线
\usepackage{siunitx}      % 数字列对齐
\usepackage{makecell}     % 可换行的表头 \makecell{...}
\usepackage{multirow}     % 跨行
\usepackage{threeparttable} % 表内脚注
\usepackage{tabularx}     % X 列型（自动拉伸）

\usepackage{etoolbox}
\usepackage{threeparttable}
\usepackage{colortbl}
\usepackage{multicol}

\usepackage{enumitem}
\setlist[itemize]{leftmargin=1.3em, topsep=2pt}

\usepackage{tikz}
\usetikzlibrary{patterns,patterns.meta}
\usepackage{pgfplots}
\usepackage{pgfplotstable}
\usepgfplotslibrary{groupplots}
\pgfplotsset{compat=1.18}
\usepackage{environ}

\usetikzlibrary{calc}
\usepackage{pgf-pie}
\usetikzlibrary{arrows.meta}
\usepackage{placeins}
\usepackage{stfloats}

\usepackage{caption}
\usepackage{subcaption}

\usepackage{marginnote}
\newcommand{\nobi}[1]{{\color{magenta}{[#1]}}}
\newcommand{\zz}[1]{{\color{orange}{[#1]}}}

\newcommand{\todo}[1]{{\color{red}{[TODO: #1]}}}

\newcommand{\inlsec}[1]{\smallskip\noindent\textbf{#1.}}

\newcommand{\sqlkw}[1]{{\color{blue}\ttfamily\bfseries\small #1}}
\newcommand{\sqlvar}[1]{{\color{black}\ttfamily\small #1}}

\algnewcommand\Input{\item[\textbf{Input:}]}
\algnewcommand\Output{\item[\textbf{Output:}]}

\newrobustcmd*{\mytriangle}[1]{\tikz{\filldraw[draw=#1,fill=#1] (0,0) --(0.2cm,0) -- (0.1cm,0.2cm);}}

\newrobustcmd*{\emptytriangle}{\tikz{\draw (0,0) --(0.2cm,0) -- (0.1cm,0.2cm) --(0,0);}}

\newrobustcmd*{\fulltriangle}{\tikz{\fill (0,0) --(0.2cm,0) -- (0.1cm,0.2cm) --(0,0);}}

\usepackage{chngcntr} % 可选，用于按 section 重置

\newcounter{rqctr}[section]             % 每个 \section 重置为 0
\newcommand{\bcirc}[1]{%
\tikz[baseline=(char.base)]{
\node[shape=circle, fill=white,draw=black,           % 画黑色圆框
  line width=0.4pt, inner sep=0.7pt] (char)
{\color{black}\bfseries\small #1};
}}

\newcolumntype{C}[1]{>{\centering\arraybackslash}p{#1}}

\makeatletter
\newsavebox{\measure@tikzpicture}
\NewEnviron{scaletikzpicturetowidth}[1]{%
  \def\tikz@width{#1}%
  \def\tikzscale{1}\begin{lrbox}{\measure@tikzpicture}%
  \BODY
  \end{lrbox}%
  \pgfmathparse{#1/\wd\measure@tikzpicture}%
  \edef\tikzscale{\pgfmathresult}%
  \BODY
}
\makeatother

\definecolor{ps1green}{HTML}{B7E6A5} % green series 1-7, greater number leads to a the thicker color
\definecolor{ps2green}{HTML}{7CCBA2}
\definecolor{ps3green}{HTML}{46AEA0}
\definecolor{ps4green}{HTML}{089099}
\definecolor{ps5green}{HTML}{00718B}
\definecolor{ps6green}{HTML}{045275}
\definecolor{ps7green}{HTML}{003147}

\definecolor{diverge1}{HTML}{045275}
\definecolor{diverge2}{HTML}{089099}
\definecolor{diverge3}{HTML}{EF8733}
\definecolor{diverge4}{HTML}{FCDE9C}
\definecolor{diverge5}{HTML}{F0746E}
\definecolor{diverge6}{HTML}{DC3977}
\definecolor{diverge7}{HTML}{7C1D6F}

\newcommand{\ourapproach}{DAT\xspace} 
\newcommand{\mytool}{DistRanger}

\definecolor{reviewblue}{RGB}{0,102,180} 
\definecolor{tagblue}{RGB}{0,120,210} 
\definecolor{r2red}{RGB}{174,45,50}
\definecolor{r3orange}{RGB}{255,112,0}
\definecolor{tagbg}{RGB}{255,255,255}

\begin{document}

\title{Distribution-Aware Distributed Database Testing \\
(Extended Version)}

\author{Zhou Zhou}
\affiliation{%
  \institution{East China Normal University}
  \country{}
  \city{}
}
%\email{52295902045@stu.ecnu.edu.cn}

\author{Si Liu}
\authornote{Si Liu (liusi@cuhk.edu.cn) and Min Zhang (zhangmin@sei.ecnu.edu.cn)
are the corresponding authors.}
\affiliation{%
  \institution{The Chinese University of Hong Kong, Shenzhen}
  \country{}
  \city{}
}
%\email{liusi@cuhk.edu.cn}

\author{Hengfeng Wei}
\affiliation{%
  \institution{Hunan University}
  \country{}
  \city{}
}
%\email{hfwei@hnu.edu.cn}

\author{Min Zhang}
\authornotemark[1]
\affiliation{%
  \institution{East China Normal University}
  \country{}
  \city{}
}
%\email{zhangmin@sei.ecnu.edu.cn}

\begin{abstract}

Distributed database management systems (DDBMSs) introduce new challenges for assessing their reliability due to distribution-specific characteristics that affect query execution and optimization. 
Existing testing approaches, largely designed for centralized DBMSs, often fail to explore diverse distributed execution behaviors 
and suffer from low executability of generated test queries, 
thereby limiting their effectiveness in bug detection.

We propose \ourapproach (Distribution-Aware Testing), a novel automated approach for detecting
query-processing bugs related to distribution strategies and distributed optimizations  in DDBMSs, by systematically leveraging distribution-aware information throughout the testing pipeline.
\ourapproach builds on a set of techniques that capture diverse combinations of logical schemas and data distribution strategies, 
and performs guided query mutation to trigger a wide range of distributed query execution behaviors and optimizations, while improving query executability via historical feedback.
We implement our approach in a tool,  \mytool, and evaluate it on four widely used 
production DDBMSs. 
It uncovers 31 previously unknown bugs, 
including 28 related to 
 distributed query processing and optimization,
and outperforms state-of-the-art testers.

\end{abstract}

%Their optimizers deploy distribution-aware techniques such as shard routing and co-located joins, but the interplay between complex SQL semantics and distributed execution makes these systems highly susceptible to subtle bugs. 

% due to distributed semantic limitations.

% \mytool's contribution is mainly about these aspects: DDBMS state generation, distributed query mutation and a feedback method.

\maketitle

\pagestyle{\vldbpagestyle}
\begingroup\small\noindent\raggedright\textbf{PVLDB Reference Format:}\\
\vldbauthors. \vldbtitle. PVLDB, \vldbvolume(\vldbissue): \vldbpages, \vldbyear.\\
\href{https://doi.org/\vldbdoi}{doi:\vldbdoi}
\endgroup
\begingroup
\renewcommand\thefootnote{}\footnote{\noindent
This work is licensed under the Creative Commons BY-NC-ND 4.0 International License. Visit \url{https://creativecommons.org/licenses/by-nc-nd/4.0/} to view a copy of this license. For any use beyond those covered by this license, obtain permission by emailing \href{mailto:info@vldb.org}{info@vldb.org}. Copyright is held by the owner/author(s). Publication rights licensed to the VLDB Endowment. \\
\raggedright Proceedings of the VLDB Endowment, Vol. \vldbvolume, No. \vldbissue\ %
ISSN 2150-8097. \\
\href{https://doi.org/\vldbdoi}{doi:\vldbdoi} \\
}\addtocounter{footnote}{-1}\endgroup

\ifdefempty{\vldbavailabilityurl}{}{
\vspace{.3cm}
\begingroup\small\noindent\raggedright\textbf{PVLDB Artifact Availability:}\\
The source code, data, and/or other artifacts have been made available at \url{\vldbavailabilityurl}. 
\endgroup
}

\section{Introduction }\label{intro}
Distributed database management systems (DDBMSs) are a key component of modern cloud computing infrastructure
and are widely adopted to scale storage and computation beyond the limits of centralized DBMSs~\cite{distrbutedsysbook}.
Unlike their centralized counterparts,
where a schema fully determines data organization,
DDBMSs additionally rely on 
\emph{distribution strategies}~\cite{dbbook},
such as partitioning and replication,
to place data across  physical nodes.
This fundamental difference significantly impacts query processing.
In particular, the execution of a SQL query in a DDBMS depends not only on its logical structure and semantics,
but also on how the involved tables are distributed.
As a result, the same query may exhibit vastly different execution behaviors,
including data access patterns and communication costs, 
under different distribution strategies.

To achieve efficient query processing under such complex execution settings, 
DDBMSs implement sophisticated distributed query optimizations,
e.g., shard routing~\cite{citus_shard_pruning} and co-located joins~\cite{shardingsphere_route_engine},
to reduce cross-node communication
and localize computation whenever possible.
However, these optimizations, as well as distributed query processing more broadly, 
are highly error-prone   
and can
result in various types of bugs and issues in practice. 
These include
logic bugs that  produce incorrect query results 
(e.g., the one shown in Figure~\ref{fig:intro-bug}),
crashes that lead to system failures,
and timeouts that prevent queries from completing within a reasonable time.
%Figure~\ref{fig:intro-bug}
%shows a logic bug in the production DDBMS Vitess~\cite{vitess}  (uncovered by our tool),
% along with the bug-triggering query,
%which persisted for nearly three years across six major versions despite extensive prior
%testing efforts.

\inlsec{Challenges in Testing DDBMSs}
Recent years have seen a surge of efforts on DBMS testing~\cite{SQLancer,DQP,EET,constant,srs,10.14778/3734839.3734861,Mozi,lego,QPG,sqlright,TQS,dynsql,10.14778/3659437.3659445}, 
which have found numerous bugs in  real-world  systems.
However, existing testing approaches are unlikely to expose bugs like the  one in Figure~\ref{fig:intro-bug}.
For example, neither of the two representative state-of-the-art testers, 
 SQLancer~\cite{SQLancer,PQS} and EET~\cite{EET},
 identified this bug in a continuous 24-hour testing campaign (see also Section~\ref{sec:Comparison}).
The underlying reason is that these approaches are largely designed for testing centralized DBMSs and are \emph{agnostic to the distribution characteristics} of DDBMSs.
Consequently, they exhibit fundamental limitations in two key aspects when applied to DDBMSs, namely \emph{test diversity} and \emph{query executability}.

First, test cases generated by existing approaches, though often complex (e.g., involving a range of SQL features and schemas),
do not account for distribution-specific factors,
such as  distribution strategies and their interactions with query semantics.
As a result, these approaches often 
 fail to capture diverse execution behaviors unique to distributed settings.
Moreover, their reliance on random testing, while effective for centralized DBMSs, often fails to trigger certain distributed query optimizations.
This is because the distributed  execution space is significantly larger and the conditions required to activate particular optimizations are rarely satisfied by chance.
For instance,  as shown in Figure~\ref{fig:intro-bug},
triggering the co-located join optimization bug in Vitess~\cite{vitess} requires specific structural patterns, i.e., two nested joins on a common sharding key \sqlvar{sdkey} (lines~12--13).
This bug persisted for nearly three years across six major versions despite extensive prior testing efforts.

Second, DDBMSs often reject SQL queries due to distribution-specific constraints
or unsupported   features~\cite{vitess_sql_support, citus_sql_support},
as such queries may require non-trivial remote execution patterns and multi-round cross-node coordination (see also Section~\ref{sec:CapabilityBoundary}).
As a result, a large fraction of queries generated by existing approaches, though valid in centralized DBMSs, fail to execute in distributed settings, which significantly limits overall testing efficacy from the outset.
For example,  we observe that,
out of 100K  queries produced by EET, 83\% are executable on PostgreSQL,
whereas only 25\% are executable on its distributed deployment Citus~\cite{citus}.

Notably, these two aspects give rise to
an \emph{inherent tension} in DDBMS testing:
increasing query complexity improves test diversity,
but simultaneously undermines executability. 
This tension is largely absent in centralized DBMS testing and poses a fundamental challenge for testing DDBMSs.

\begin{figure}[t]
\centering
\begin{minipage}{.97\linewidth}

\begin{lstlisting}[style=sql, firstnumber=1]
CREATE TABLE t0 (sdkey INT);
CREATE TABLE t1 (sdkey INT);
\end{lstlisting}

\begin{lstlisting}[style=json, firstnumber=last]
/* VSchema (specify dist. strategies per table) */
{"sharded": true, "tables": {
    "t0": { "column_vindexes": [{ "column": "sdkey", "name": "i" }] },
    "t1": { "column_vindexes": [{ "column": "sdkey", "name": "i" }] } },
  "vindexes": { "i": { "type": "xxhash" } } }
\end{lstlisting}

\begin{lstlisting}[style=sql, firstnumber=last]
INSERT INTO t0 VALUES (NULL);
INSERT INTO t1 VALUES (1);

SELECT * FROM (t0 AS r0 LEFT OUTER JOIN
  (t1 AS r1 JOIN t1 AS r2 ON (r1.sdkey = r2.sdkey))
    ON (r0.sdkey = r1.sdkey)); 
 --- expected result: one row returned
 --- actual result: none (*@\faBug@*) 
\end{lstlisting}

\end{minipage}
 \captionsetup{skip=8pt}
\caption{A logic bug in Vitess found  by our tool, along with the test query. 
Vitess incorrectly returns an empty result set, while the expected result contains one row.
A JSON-based VSchema specifies distribution strategies per table 
(lines 4--7). \sqlvar{sdkey} denotes the common sharding key of the two tables.
}
\label{fig:intro-bug}
%\vspace{-2ex}
\end{figure}

\inlsec{Our Approach}
We introduce \ourapproach (Distribution-Aware Testing), a novel automated testing approach for DDBMSs. 
It effectively discovers 
query-processing bugs related to distribution strategies and distributed optimizations 
that are unlikely to be uncovered by existing  testing techniques.
Our key insight is that distribution-aware information, 
i.e., data distribution strategies and their interaction with query semantics, 
can be systematically leveraged throughout the testing pipeline to address the aforementioned challenges.

\ourapproach
begins by constructing a \emph{Schema-Distribution Matrix} (SDM), 
which captures a wide range of combinations across diverse logical schemas and data distribution strategies. 
This provides a structured yet expressive exploration space for generating varied DDBMS instances for testing.
In particular,
this design offers two key properties:
(i) column-wise, tables derived from the same schema can be interchanged to explore different distribution strategies without violating SQL syntactic validity; and
(ii) row-wise, tables instantiated under aligned distribution strategies naturally satisfy the conditions required by certain distributed query optimizations.

These properties enable effective test query generation over SDM, 
which \ourapproach achieves through a generic mutator $\mu = (\rho, \Delta)$. 
Specifically, in addition to column-wise and row-wise table substitutions via $\rho$,
it incorporates query transformations encoded in $\Delta$, tailored to specific distributed query optimizations, such as shard routing and co-located joins. 
By applying sequences of mutations using $\mu$,
\ourapproach-generated queries can trigger a range of distributed  execution behaviors and optimizations.

To improve query executability, 
\ourapproach incorporates a feedback mechanism during query mutation.
By leveraging the history of previously executed and rejected queries in the DDBMS under test,
it learns patterns of unsupported query structures under specific distribution strategies,
which are encoded as \emph{feature paths}, 
and guides subsequent generation toward diverse yet executable queries. 
This feedback loop, in turn,
effectively mitigates the tension between test diversity and query executability.

We implement \ourapproach in a tool called \mytool. 
It currently employs differential testing as the primary test oracle, 
i.e., identifying discrepancies between the execution results of the same query under centralized and distributed deployments of the same DBMS; any discrepancy indicates potential issues specific to the distributed layer. 
However, \ourapproach can naturally incorporate other types of oracles, e.g., metamorphic ones~\cite{TLP,EET}.

In addition, \ourapproach can seamlessly integrate advances in query and schema generation into base query generation prior to mutation and SDM construction, respectively. 
Finally, the mutator $\mu$ is general and can be instantiated to incorporate additional distributed query optimizations beyond our current focus on shard routing and co-located joins. 
Together, these techniques establish \ourapproach as a general framework for testing distributed query processing and optimization in DDBMSs.
Sections~\ref{sec:related} and~\ref{sec:concl} 
provide a detailed discussion of these aspects.

\inlsec{Contributions} Overall, we make the following contributions.

\begin{itemize}
    \item
Conceptually, we propose \ourapproach, a novel approach for detecting query-processing bugs in  DDBMSs that are related to distribution strategies and distributed  optimizations.
This problem presents fundamental challenges beyond centralized DBMS testing, which we address by leveraging distribution-aware information throughout the testing pipeline.
       
%a novel automated approach for testing  DDBMSs by systematically leveraging distribution-aware information throughout the testing pipeline. 
    
    \item  Technically, we develop a set of techniques that generate diverse test cases, including DDBMS instances and distributed queries, improve query executability, and mitigate the inherent tension between diversity and executability.
    
    \item Practically, we realize \ourapproach in an automated tester, \mytool, and assess  it on four production DDBMSs: Citus, Vitess, Apache ShardingSphere,  and ClickHouse.
It uncovers 31 previously unknown bugs (21 of which have been confirmed and 15 of which have been fixed), including 28 bugs whose manifestations or root causes are tied to distributed query processing and optimization.
Compared to the state-of-the-art testers, \mytool{} demonstrates superior effectiveness in 
 %both the generation of diverse yet executable distributed queries and
 bug detection. 
\end{itemize}

\section{Background }\label{sec:background}
%Despite the great diversity of DDBMSs, techniques used in \mytool{} are based on a few common natures of them.
% To make clear terminologies for better describing \mytool, we have to give some detailed definition of DDBMS related to our testing steps.
%This section provides important background information on data-distribution strategies, optimization and limitations of DDBMSs.

%\begin{figure*}[t]
%    \centering
%    \includegraphics[width=\textwidth]{figures/dds types.pdf}
%      \captionsetup{skip=5pt}
%    \caption{
    %A unified terminology for distribution strategies (DSs) and
%    DDBMS-specific specifications.}
%    \label{fig:ddsTypes}
%\end{figure*}

\subsection{Distribution Strategies} \label{subsec:ds}
% In DDBMSs, the physical distribution of data directly affects both the parallel execution capability of queries and the system's availability. 

In a centralized DBMS, a schema defines the logical organization of data as a set of relations (or tables).
Yet, in a DDBMS, the schema alone does not determine data placement. 
The system must additionally define 
\emph{distribution strategies},
which specify how data is physically partitioned for scalability or replicated for availability across nodes~\cite{dbbook}. 
These strategies are typically defined \emph{per table}.
%Although DDBMSs 
%such as
%Citus~\cite{citus},
% ShardingSphere~\cite{shardingsphere},   Vitess~\cite{vitess},
%and ClickHouse~\cite{clickhouse}
%use varying terminology for similar concepts, we adopt a unified set of terms to avoid ambiguity:
%\textit{local tables}, \textit{reference tables}, \textit{distributed tables}, and \textit{co-located tables}.\footnote{\todo{...}}  

In practice, DDBMSs often use different terminology for conceptually similar distribution strategies; 
Table~\ref{tab:dds-terms} summarizes the terminology across four representative systems.
To avoid ambiguity, we adopt the terminology of Citus~\cite{citus} in this paper.

\begin{description}[leftmargin=10pt]
    \item[Local tables] store all data on a single node, allowing queries over them to be executed without cross-node communication.

    \item[Reference tables] 
   store a full copy of the table on each node,
   either via explicit replication (e.g., in Vitess)  or router-level broadcast (e.g., in ShardingSphere).
   This enables local access during distributed query execution.

\item[Distributed tables]
are horizontally partitioned across  nodes based on a sharding key and sharding method.
Querying such tables requires accessing multiple nodes and merging partial results.

\item[Co-located tables] 
are distributed tables with aligned partitioning schemes, such that 
rows sharing the same sharding key reside on the same node. 
This enables efficient joins on the sharding key, commonly referred to as \emph{co-located joins}~\cite{shardingsphere_route_engine}.

\end{description}

\begin{table}[t]
  \captionsetup{skip=5pt}
  \caption{Terminology for distribution strategies across four  DDBMSs. Each row corresponds to a distribution strategy. 
  %This paper adopts  the terminology of Citus.
  }
  \label{tab:dds-terms}
  \centering
  \resizebox{\columnwidth}{!}{
  \begin{tabular}{l|lll}
    \toprule
     \textbf{Citus} & \textbf{ShardingSphere} & \textbf{Vitess} & \textbf{ClickHouse} \\
    \midrule
\rowcolor{gray!5}     Local~\cite{citus_table_types}
      & Single~\cite{ss_single_table_rule}
      & Unsharded~\cite{vitess_unsharded}
      & MergeTree~\cite{clickhouse_mergetree} \\
   \rowcolor{gray!15}  Reference~\cite{citus_table_types}
      & Broadcast~\cite{ss_broadcast_table_rule}
      & Reference~\cite{vitess_reference_tables}
      & %Replicated~\cite{clickhouse_replication}
      N/A\\
 \rowcolor{gray!25}    Distributed~\cite{citus_table_types}
      & Sharding~\cite{ss_sharding_table_rule}
      & Sharded~\cite{vitess_sharded}
      & Distributed~\cite{clickhouse_distributed} \\
\rowcolor{gray!35}     Co-located~\cite{citus_colocation}
      & Binding~\cite{ss_sharding_table_reference_rule}
      & Same Vindex~\cite{vitess_lookup_primary}
      & N/A \\
    \bottomrule
  \end{tabular}
  }
\end{table}

Different DDBMSs often specify these distribution strategies using different languages, e.g., SQL or SQL-like DDL (Data Definition Language) statements in Citus, ShardingSphere, and ClickHouse, and a JSON-based schema in Vitess (Figure~\ref{fig:intro-bug}).
Note that DDBMSs may impose restrictions on certain strategies (e.g., in Citus, local tables exist only on the coordinator node), and some systems provide additional system-specific variants~\cite{tidb_tiflash_overview}.
A table may also simultaneously exhibit both partitioning and replication (e.g., replicated shards); yet, such cases can still be expressed using these strategies.

\subsection{Distributed Query Processing and
Optimization}\label{sec:Optimization}
%In DDBMSs, a query is processed by decomposing it into fragments, executing them across nodes, and coordinating multiple rounds to produce the final result.
%In this process, 
%\emph{distributed query optimization} plays a critical role in system performance by determining execution strategies that minimize  cross-node communication overhead and localize computation whenever possible.
In a DDBMS, data is often partitioned across multiple nodes (or \emph{shards}), 
and a query may need to access one or multiple shards depending on the data it touches. 
Hence, executing a query involves not only computing query results but also coordinating data access and communication across nodes.
Distributed query optimization aims to minimize such communication overhead by exploiting data distribution and choosing efficient execution strategies.

%Unlike single-node DBMS, DDBMSs may parse the query into several parts, distribute them to physical nodes and coordinate a few rounds to collect the final result.
%A unifying principle of DDBMS optimization is to reduce communications or localize remote work in advance. 

Below,
we introduce two representative distributed query optimizations,
 \emph{shard routing} and \emph{co-located joins},
which are widely supported by existing DDBMSs~\cite{citus_distributed_readme,shardingsphere_route_engine,vitess_query_plan_metrics}.
% and also serve as running examples throughout the paper.
 %\todo{justify why these two}

 % and often serve as prerequisites for broader downstream optimizations

%To satisfy some special DDBMS optimizations, we need not only structure-static preconditions in SQL query, but also precisely DDS-related structures. 
%We call them \emph{Distribution-Sensitive Optimizations}. We introduce shard routing and co-located join as two commonly-used distribution-sensitive optimization primitives. They are often exposed as first-class optimizations in DDBMSs and serve as prerequisites for broader downstream optimizations~\cite{citus_distributed_readme,shardingsphere_route_engine,vitess_query_plan_metrics}.

\begin{figure}[t]
  \centering
  \includegraphics[width=\columnwidth]{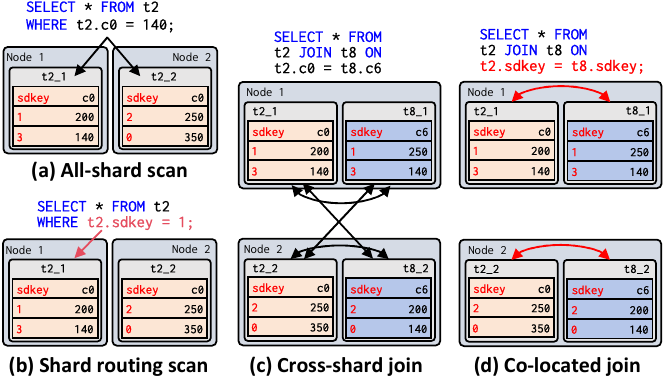}
  \captionsetup{skip=5pt}
  \caption{An illustration of the two distributed query optimizations: shard routing (a vs. b) and co-located joins (c vs. d). 
  In each case, both tables are sharded by \sqlvar{sdkey}.
}
  \label{fig:optimization}

\end{figure}

\begin{figure}[t]
  \centering
  \includegraphics[width=\columnwidth]{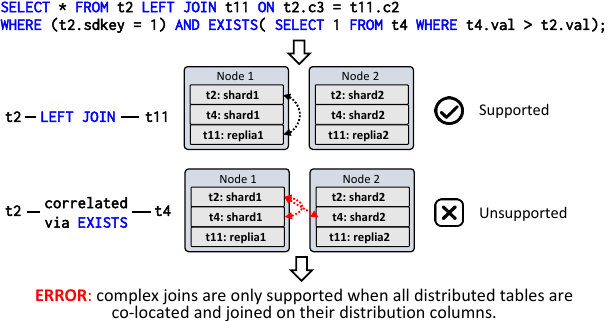}
  \captionsetup{skip=8pt}
  \caption{An example of a  SQL query rejected by Citus (v13.0).}
  \label{fig:limitation}
  \vspace{-1ex}
\end{figure}

\inlsec{Shard Routing} 
Since a distributed table is partitioned into multiple shards, scanning the entire table may require accessing all shards,
which is referred to as \emph{all-shard access}.
However, if a query predicate identifies the shard containing the requested data, the DDBMS can avoid unnecessary accesses.
%Scanning a distributed table often requires accessing all of its shards.
%However, when a query includes a predicate on the sharding key, execution can be restricted to the shards that satisfy the predicate. 
This optimization is known as \emph{shard routing} (or shard pruning)~\cite{citus_shard_pruning,shardingsphere_route_engine}.

% \zz{old: DDBMSs would check if any parts of the query can be pushed down to only a few nodes rather than every shard-holding node. This optimization works by reducing the number of shards that need to be accessed~\cite{citus_shard_pruning,shardingsphere_route_engine}.This optimization could be performed on any distributed table \sqlvar{t}. Its DDS-related structures is a predicate on sharding key of \sqlvar{t}. Thus DDBMSs only need to interact with shards that satisfy this predicate.}

\begin{example}
In Figure~\ref{fig:optimization}(a) and (b),
 the table \sqlvar{t2} is sharded by 
\sqlvar{t2.sdkey mod 2}.
As a result, querying \sqlvar{t2}
  often requires an all-shard scan 
  accessing both nodes,
  e.g., by
  \sqlkw{SELECT} \sqlvar{$\ast$} \sqlkw{FROM} \sqlvar{t2}  \sqlkw{WHERE} \sqlvar{t2.c0 = 140}
(Figure~\ref{fig:optimization}(a)).
  However, when the \sqlkw{WHERE} clause includes the predicate \sqlvar{t2.sdkey = 1}, the DDBMS can leverage the sharding strategy of \sqlvar{t2} to route the query to Node~1 only, as shown in Figure~\ref{fig:optimization}(b).

\end{example}

% In general, if a selection will be filtered by a predicate \(\texttt{t1.id} \equiv N \pmod{2}\), where \(K\) is a sharding key of \(R\), and \(K^*\wedge(p_{r_i}\vee,\cdots,\vee p_{r_j})\) is a tautology, this selection only needs data from nodes \(n_i\,\cdots,\ n_j\) that hold necessary shards. A shard router then push the selection down to those relevant nodes, preventing full-table scanning from massive nodes. In Figure~\ref{fig:optimization}(a), the selection could only push down to one node if its \emph{where clause} contains predicate \(\texttt{t2.id = 3}\).

\begin{figure*}[t]
  \centering
  \includegraphics[width=\textwidth]{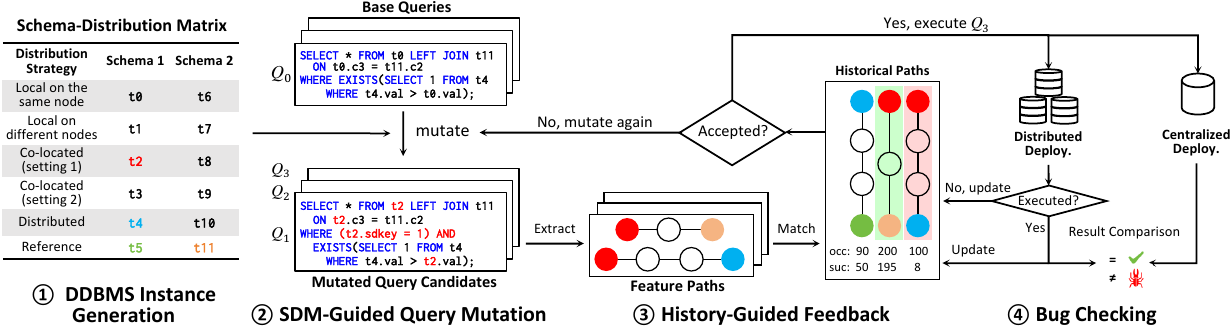}
    \captionsetup{skip=8pt}
    \caption{Workflow of our \ourapproach approach. }
    \label{fig:overview}
\end{figure*}

\inlsec{Co-located Joins} 
%A join between two distributed tables often requires cross-shard communication.
%However, if the tables are co-located, a join on the sharding key can be executed locally, as rows with the same sharding key are placed on the same node. 
A join combines rows from two tables based on a join condition.
In a DDBMS, joining two distributed tables may require accessing rows across multiple shards  
when matching rows cannot be found locally.
Such joins are referred to as \emph{cross-shard joins} and can result in significant communication overhead.
However,
if the tables are co-located, 
i.e., rows with the same sharding key are placed on the same nodes, 
a join 
 on the sharding key
 can be executed locally
  without accessing remote nodes. 
This optimization is known as  \emph{co-located joins}.

\begin{example}
In Figure~\ref{fig:optimization}(c) and (d), %the tables 
\sqlvar{t2} and \sqlvar{t8} are co-located on \sqlvar{sdkey}, meaning that rows with the same \sqlvar{sdkey} reside on the same node. In Figure~\ref{fig:optimization}(d), the join condition \sqlvar{t2.sdkey = t8.sdkey} allows each shard of \sqlvar{t2} to join only with the corresponding local shard of \sqlvar{t8}.
%without accessing data from remote nodes.
In contrast, in Figure~\ref{fig:optimization}(c), the join is on \sqlvar{c0} and \sqlvar{c6} rather than the sharding key. 
As a result, each shard of \sqlvar{t2} may need to join with all shards of \sqlvar{t8}, 
potentially
incurring
high communication overhead.

%In Figure~\ref{fig:optimization} (c) and (d), both tables \sqlvar{t2} and \sqlvar{t8} are already co-located by \sqlvar{sdkey}, meaning that rows of these two tables with the same \sqlvar{sdkey} value are placed on the same node. In Figure~\ref{fig:optimization} (d), the join condition is \sqlvar{t2.sdkey = t8.sdkey}. Therefore, on Node 1, the shard of \sqlvar{t2} only needs to join with the local shard of \sqlvar{t8}, without accessing data from other nodes. In contrast, in Figure~\ref{fig:optimization} (c), the join condition is on the \sqlvar{c0} and \sqlvar{c6} column rather than the sharding key. As a result, each shard of \sqlvar{t2} may need to join with every shard of \sqlvar{t8}. Since these shards are distributed across different nodes, the join may require cross-node communication.
\end{example}

%In addition to shard routing and co-located joins, DDBMSs may implement other distributed query optimizations.
%Nevertheless,
% a common theme is that these optimizations leverage data distribution strategies to reduce query execution cost  in distributed settings.

% modern DDBMSs apply a variety of optimizations that are largely orthogonal to DDS patterns and are typically triggered automatically by the optimizer. These techniques mainly improve intra-node work or constant factors. In contrast, \emph{shard routing} and \emph{co-located joins} determine whether execution can be kept local and thus qualitatively change the cross-shard communication pattern (and even feasibility). Hence our generator targets these two mechanisms explicitly, while the remaining optimizations are naturally covered by general workload synthesis.

% Some DDBMSs may support \emph{distributed query plan} to assess how well a SQL query is optimized. A well-optimized plan executes most work on the shards that own the relevant data, rather than transmitting all partitions to the coordinator for centralized processing. In plan terms, this shows up as more operators pushed down to data nodes and fewer communication steps. For a testing tool, a primary objective is to generate workloads that induce a diverse set of distributed plan shapes, thereby exercising different optimization paths.

\subsection{Unsupported SQL Features in DDBMSs}\label{sec:CapabilityBoundary}

Although distributed query optimization improves execution efficiency, 
DDBMSs often conservatively reject complex SQL queries at planning time and return errors indicating \emph{unsupported features}~\cite{vitess_sql_support, citus_sql_support}.
This is because processing such queries may introduce non-trivial remote execution patterns, and some executions may require multiple rounds of coordination across nodes, 
which fall outside the system's design assumptions.

Note that the set of unsupported SQL features may vary across DDBMSs and even across different versions of the same system. 
Nevertheless, these limitations are commonly 
rooted in mismatches between query semantics and data distribution constraints.

The following  illustrates a SQL query that is not executable in Citus (v13.0)
and serves as a running example throughout the paper.

%While DDBMSs provide scalability and parallelism, their query processing may introduce complex remote execution. 
%Some execution may require coordination across nodes for several times, which may not be supported by a specific DDBMS.
%Thus, DDBMS may reject such a complex SQL at planning time and returns an error indicating that the query contains unsupported features~\cite{vitess_sql_support, citus_sql_support}. 
% When distribution-aware optimizations are inapplicable, structures such as \emph{cross-shard joins} and \emph{nested correlated subqueries} induce overwhelming communication cost among nodes.
% In addition, for update-heavy DML, maintaining the all shards and replicas may further trigger large-scale data migration across nodes. 
% In DDBMSs especially that were not designed from scratch, the coordinator has limited control over data nodes, so that communication patterns of such plans are typically deemed impractical and therefore rejected.

\begin{example}

As shown in Figure~\ref{fig:limitation}, 
\sqlvar{t2} and \sqlvar{t4} are two distributed tables,  
while \sqlvar{t11} is a reference table.
The  query involves
a shard-routing predicate \sqlvar{t2.sdkey = 1} that restricts access  to a shard of \sqlvar{t2},
as well as
two types of interactions between tables.  
The first arises from  \sqlvar{t2} \sqlkw{LEFT JOIN} \sqlvar{t11}, which is supported by Citus.
The second is an \sqlkw{EXISTS}-based correlation between \sqlvar{t4} and \sqlvar{t2}. 
Despite shard routing on \sqlvar{t2}, the correlation with \sqlvar{t4} still induces a cross-shard semi-join, which is not supported. 
As a result, the planner fails to produce a query plan for the entire query, which is ultimately rejected by Citus with an error.
%indicating complex joins are only supported when all distributed tables are co-located and joined on their distribution columns.

Query executability poses a major obstacle to DDBMS testing: 
a high rejection rate of generated queries prevents effective exploration of system behavior, 
thereby limiting bug detection.

%Figure~\ref{fig:limitation} shows a query rejected by Citus 13.0. Here, \sqlvar{t2} and \sqlvar{t4} are randomly generated distributed tables, where \sqlvar{t2} has a shard routing predicate \sqlvar{t2.sdkey = 1}, and \sqlvar{t11} is a reference table. 
%The query contains two structures shown as two relationships between tables. The first relationship comes from \sqlvar{t2} \sqlkw{LEFT JOIN} \sqlvar{t11}, which is supported by Citus. 
% \zz{remove: Since \sqlvar{t11} is a reference table, every shard of \sqlvar{t2} can find a complete local copy of \sqlvar{t11}, so this part can be executed as a local join. This structure is therefore valid and accepted by the planner.} 
%The second relationship is the \sqlkw{EXISTS}-based correlation between \sqlvar{t4} and \sqlvar{t2}. Unfortunately, even with the help of shard routing, it still requires a cross-shard semi-join, which is not supported. As a result, the planner refuses to produce a query plan for this structure. 
%Thus, the DDBMS eventually rejects the entire query with an error. 
\end{example}

%These unsupported features may differ across different DDBMSs and different versions of a DDBMS~\cite{citus_13_release_notes}, but they are commonly caused by unsupported semantics of a specific relationship.

% DDBMSs--especially those that were not designed from scratch--often has only limited control over data nodes. As a result, such systems tend to exhibit more distribution-related query limitations. Moreover, these limitations can also be rejected by extra statement validation or DDL/DML checking instead of query planner.
% To make things worse, each DDBMS makes product-specific trade-offs: some complex constructs are supported while others are not, and the limitations evolve across versions. A common feature of such limitations is that they are triggered by a specific syntactic structure together with particular DDS semantics, and are therefore distribution-related.

% In practice, queries are easy to be invalid when testing tools try to generate complex structures by extending the height of abstract syntax trees or introducing advanced SQL features. 
% For example, SQLancer's~\cite{SQLancer} validity is only 32\% for testing Citus~\cite{citus}, despite limit generation on complex structures due to its test oracles.
% Thus, overcoming the trade-off between validity and complexity has become a main problem for testing DDBMSs.

\section{Overview of  \ourapproach 
}\label{sec:Approach}
This section  provides an overview of \ourapproach (Distribution-Aware Testing), an automated  approach for 
detecting query-processing bugs in  DDBMSs that are related to distribution strategies and distributed  optimizations.
\ourapproach is unique in its ability to generate diverse yet executable queries that can effectively exercise a wide range of distributed  execution behaviors and optimizations, 
thereby uncovering issues 
specific to 
the distributed layer of DDBMSs.
%that do not arise in centralized DBMSs.

Figure~\ref{fig:overview} shows \ourapproach's workflow, consisting of four main steps.

%In this section, we present
%\ourapproach (Distribution-Aware Testing), an automatic testing approach for detecting bugs in DDBMSs.
%\ourapproach is unique in ability to generate queries explicitly triggering distributed optimizations with high validity.   
% \ourapproach is unique in its ability to generate DDBMS states and SQL queries that trigger distribution-related optimizations while maintain high validity and complexity against distribution-related limitations. 
%\subsection{Overview}

\inlsec{\bcirc{1} DDBMS Instance Generation}
\ourapproach starts by generating a set of logical schemas, including tables and their relationships (e.g., via foreign keys), and populating them with random data.
 It then generates a set of distribution strategies, each with multiple instances depending on how data is placed.
Together, the schemas and distribution strategies form  the \emph{schema–distribution matrix} (SDM), 
%which underlies the testing of individual DDBMS instances.
which
provides a structured yet expressive space for generating diverse DDBMS instances 
and guiding query mutation in Step~\bcirc{2}.

%Moreover, for each schema, 
%the SDM spans all considered distribution strategies,
%yielding a rich mutation space for the next step. 
%See Section~\ref{subsec:matrix} for details.

Figure~\ref{fig:overview} shows an example SDM containing two logical schemas and six distribution strategies, 
including randomly distributed tables, reference tables, two types of co-location settings, 
and tables localized on a single node or across different nodes.
Overall, this yields 12 distributed table instances.

\inlsec{\bcirc{2} SDM-Guided Query Mutation}
Given a DDBMS instance, \ourapproach generates a base query and applies mutations to diversify it, aiming to trigger a range of distributed execution behaviors and optimizations. 
In particular, these mutations are guided by the SDM,
e.g., by replacing a table with another one under a different distribution strategy to explicitly trigger the corresponding optimizations.

%To directly generate queries that trigger distributed query optimizations, \ourapproach adopts optimization oriented query mutation.
%Given a query, \ourapproach leverages distribution strategies to either introduce a shard-routing predicate or transform a join into a co-located join, so as to directly trigger the corresponding optimization.

%As illustrated in Figure~\ref{fig:overview}, \ourapproach typically selects one table from the original SQL query, for example the local table \sqlvar{t0}, on which shard routing cannot be applied directly.
%It then replaces \sqlvar{t0} with another table from the same column of the SDM, such as \sqlvar{t2}, and inserts a shard-routing predicate \sqlvar{t2.dkey = 1} into the \sqlkw{WHERE} clause.
%As a result, the mutated query is much more likely to trigger shard routing in the target DDBMS.

Figure~\ref{fig:overview} shows an example of triggering the shard-routing optimization.
Specifically, we select a table from the base query $Q_0$, namely the local table \sqlvar{t0}, on which shard routing cannot be directly applied.
We then replace \sqlvar{t0} with another table from the same column of the SDM, namely \sqlvar{t2}, and insert a shard-routing predicate \sqlvar{t2.sdkey = 1} into the \sqlkw{WHERE} clause.
This mutation increases the likelihood of triggering shard routing in the DDBMS under test.

Additional mutation types are described in Section~\ref{subsec:mutation}.

%In Figure~\ref{fig:overview}, this mutation is summarized as $(t0, (t0 \mapsto t2), SRH)$, where SRH stands for the mutation name \zz{\emph{Shard-Routing Hint}}.

%Besides SRH, \ourapproach also supports a co-located-join-oriented mutation called \emph{Colocation Closure} (CCC), as well as a simpler mutation called \emph{Syntax-Preserving Replacement} (SPR), which changes only table names without further modifying the SQL structure.
%As we discuss later, this mutation can potentially support a broader range of distributed query optimizations.

% Starting from a valid original query, \ourapproach performs mutation by replacing selected table references with other instances from the same \zz{SDM} column, such as using \sqlvar{t2} to replace \sqlvar{t0}. 
% This replacement is syntactically valid, but it changes the distributed semantics of the query because of DDS changing. 
% As a result, the mutated query may trigger a different distributed execution path. 

% In addition to table replacement, \ourapproach also uses distribution strategies to edit queries and explicitly steers them towards \zz{distributed query optimizations}. For example, the mutation in Figure~\ref{fig:overview} adds a shard routing predicate \sqlvar{t2.dkey = 1}.

\inlsec{\bcirc{3} History-Guided Feedback} 
Mutated queries may be rejected by a DDBMS due to  unsupported features (Section~\ref{sec:CapabilityBoundary}), which in turn reduces overall testing efficacy. 
To address this, \ourapproach incorporates a feedback mechanism that interacts with Step~\bcirc{2}, leveraging the history of previously executed and rejected queries
by the DDBMS 
to guide generation toward executable queries.

Specifically, this mechanism extracts structural features from a mutated query,
e.g., join types and correlations,
and encodes them as \emph{feature paths} 
(intuitively, how tables are related through SQL operations). 
These paths are then matched against historical paths, which also maintain executability statistics (e.g., their frequency of occurrence and successful execution counts).

%Specifically,
%this mechanism operates by extracting structural features from a mutated query,
%e.g., join types and correlations,
%and encoding them as  \emph{feature paths}
%(intuitively, how two tables are  %syntactically 
%related through SQL operations). 
%These paths are then matched against historical paths,
%which also maintain statistics on the executability of individual paths observed so far (e.g., their frequency of occurrence and the number of times they were successfully executed by the DDBMS under test).

As shown  in Figure~\ref{fig:overview}, the mechanism extracts structural features from a mutated query $Q_1$,
such as the correlation between \sqlvar{t2} and \sqlvar{t4} and the presence of the \sqlvar{t2} \sqlkw{LEFT JOIN} \sqlvar{t11} construct,
and encodes them as two feature paths.
\ourapproach rejects this candidate because one  path matches a frequently rejected pattern in the historical record (in \colorbox{red!20}{red}),
despite another having a high success rate
(in \colorbox{green!20}{green}).
As mutation proceeds iteratively, a subsequent candidate $Q_3$ 
 that passes this check is submitted to the DDBMS for execution.

\inlsec{\bcirc{4} Bug Checking}
%In this step,
\ourapproach  executes the mutated query against the DDBMS under test to detect bugs.
As shown  in Figure~\ref{fig:overview}, 
it employs differential testing to identify discrepancies between the execution results of the same query (e.g., $Q_3$) under centralized and distributed deployments of the same system; 
any mismatch indicates a
potential  
query-processing bug 
%related to distribution strategies and distributed optimizations 
related to the distributed layer.
%\ourapproach also naturally supports other test oracles; see Sections~\ref{sec:DiscoverBugs} and~\ref{sec:related}.

Note that, regardless of whether a query is successfully executed, 
this information is fed back to Step~\textbf{\bcirc{3}} to guide future mutations.

The above  steps illustrate one iteration of our testing process, 
which can be repeated. 
In subsequent iterations, we can either continue from Step~\textbf{\bcirc{2}}
with a new base query 
or restart from Step~\textbf{\bcirc{1}} to explore a new DDBMS instance with different SDMs.
Step~\textbf{\bcirc{3}} persists throughout the entire testing process,
recording information about all previously executed or attempted queries.
%and thereby improving overall testing efficacy.
%Before testing begins, 
%we configure both the number of queries generated per DDBMS instance, \todo{and more?}

%The above four steps illustrate one iteration of our testing process, which can be repeated. In subsequent iteration, we can either proceed with Step \textbf{\bcirc{2}} to generate a new query and mutate it for bug detection, or restart with Step \textbf{\bcirc{1}} to explore a new DDBMS instance. Step \textbf{\bcirc{3}} will last over the whole testing process, recording all queries and enhancing overall test validity. Before testing begins, we configure both the number of generated queries per generated DDBMS instances.

\ourapproach leverages recent advances~\cite{SQLsmith,dynsql} in centralized DBMS testing to produce syntactically valid  SQL queries and  diverse schemas; these are not part of our contributions.
The next section presents \ourapproach's core techniques tailored for testing DDBMSs.

\section{Core Techniques in \ourapproach }

% \begin{table*}[t]
% \caption{Mutation rules used in \mytool.}
% \label{tab:mutation-rules}
% \centering
% \small
% \begin{tabular}{c l l l l}
% \toprule
% \textbf{Type} & \textbf{Name} & \textbf{Precondition} & \textbf{Transformation Rule} \\
% \midrule

% base
% & Table Replacement
% & None
% & $table(\mathcal{M}[d,p]) \;\rightarrow\; table(\mathcal{M}[d',p])$ \\
% \midrule
% optional
% & Shard-Routing Predicate
% & distributed table \sqlvar{t} sharded on \sqlvar{dkey}
% & $where\_clause \rightarrow $\sqlkw{WHERE} \sqlvar{t.dkey}$=\mathit{const\_expr} \;$\sqlkw{AND}$\; \mathit{bool\_expr}$ \\

% optional
% & Co-location Rewriting
% & co-located joined tables on \sqlvar{ckey}
% & $\mathit{join\_cond} \;\rightarrow\; $\sqlkw{ON}$\; $\sqlvar{left.ckey}$=$\sqlvar{right.ckey} \\

% \bottomrule
% \end{tabular}
% \end{table*}

\begin{figure}[t]
  \centering
  \includegraphics[width=\columnwidth]{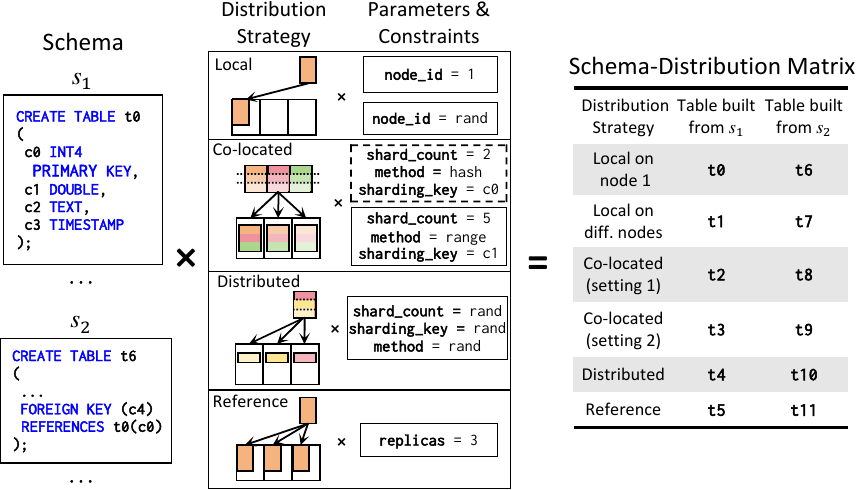}
   \captionsetup{skip=5pt}
  \caption{An illustration of how an SDM is constructed.
}
  \label{fig:sdm}
\end{figure}

\subsection{DDBMS Instance Generation} \label{subsec:matrix}

When generating a database instance for testing, unlike in the centralized setting where only the logical schema  is considered, 
the distributed setting must additionally account for how tables are partitioned or replicated across physical nodes.
To this end, we organize table instances in a DDBMS using the \emph{Schema-Distribution Matrix} (SDM), denoted by $\mathcal{M}$.
Let $\mathcal{S}$ be a set of  logical schemas
and $\mathcal{D}$ be a set of distribution strategies under consideration, with $|\mathcal{S}| = n$ and $|\mathcal{D}| = m$. 
Then $\mathcal{M}$ is an $m \times n$ matrix indexed by $\mathcal{D} \times \mathcal{S}$, where each entry $\mathcal{M}[d, s]$ denotes the table instance obtained by instantiating schema $s \in \mathcal{S}$ under distribution strategy $d \in \mathcal{D}$ 
and populating it with randomly generated data.

Figure~\ref{fig:sdm} illustrates the construction of the SDM shown in Figure~\ref{fig:overview}, which consists of two schemas and six distribution strategies.
First, we generate a set of  schemas, e.g., $s_1$, which specifies all information required to define a table in a centralized setting, including its attributes, data types,
and integrity constraints.
Next, each schema, instantiated with random data, is combined with a distribution strategy to form a distributed table instance.
For example, $s_1$ is paired with six different distribution strategies to generate the table instances \sqlvar{t0} through \sqlvar{t5}.

How are these distribution strategies generated? 
By default, we randomly configure their parameters, such as the number of shards, the sharding method, and the sharding key. 
This yields a distributed table instance in the ``Distributed'' row of the SDM, e.g., \sqlvar{t4}.

While random generation is generally effective for exploring the space of distribution configurations,
it  may be insufficient for deliberately triggering certain distributed query execution logic and optimizations, which often require multiple distributed tables to satisfy alignment conditions.
For example, co-located joins require participating tables to share the same sharding parameters.
To this end, 
\ourapproach introduces \emph{constraints}
on  distribution parameters.

\begin{example}
Consider the ``Co-located'' case, specifically the dashed box, in Figure~\ref{fig:sdm}.
\ourapproach defines a parameter setting for co-located tables by constraining the shard count (2), the sharding method (hash), and the sharding key (\sqlvar{c0}).
It then instantiates multiple tables under this setting.
As a result, table instances in the same row ``Co-located (setting 1)'', e.g., \sqlvar{t2} and \sqlvar{t8}, are naturally aligned and thus co-located, as they share the same sharding parameter setting, even though they are instantiated from different schemas.

%These settings correspond to separate rows in the SDM, e.g., ``Co-located (setting 1)'' and ``Co-located (setting 2)''.
\end{example}

When generating an SDM, 
 although the types of distribution strategies (Table~\ref{tab:dds-terms}) are relatively limited compared to schemas,
we consider a larger number of their instantiations to enable diverse distributed execution behaviors and optimizations. 
The exact number
$|\mathcal{D}|$
 is determined by the testing goal and the alignment conditions required to trigger specific optimizations, and may vary  across DDBMSs due to differences in  their supported  strategies.

\begin{example}

As shown in Figure~\ref{fig:sdm}, \ourapproach introduces two parameter constraints for local tables, and similarly for co-located tables. 
This results in two distinct rows in each case.
 For example,  
 local tables may be placed on a fixed node (both \sqlvar{t0} and \sqlvar{t6} on Node 1) or on  different nodes (for \sqlvar{t1} and \sqlvar{t7}).
Compared to 
``Co-located (setting 1)'',
\sqlvar{t3}  and \sqlvar{t9} 
are co-located tables instantiated under a different parameter setting (five shards, range sharding on the key \sqlvar{c1}).
\end{example}

%Note that
% the SDM not only provides a strong basis for generating diverse DDBMS instances, but also guides query mutation to explore a broader range of distributed query execution logic and optimizations,
% as we will see next.

 Note that
in an SDM, tables %such as \sqlvar{t0} and \sqlvar{t2} 
share \emph{the same schema column-wise}.
Replacing one with the other preserves SQL syntactic validity while potentially inducing 
different distributed execution behaviors.
Tables %such as \sqlvar{t2} and \sqlvar{t8}
share \emph{the same distribution strategy row-wise}, enabling preconditions required by certain optimizations (e.g., co-located joins). 
Overall, the SDM  provides a strong basis not only for generating diverse DDBMS instances, 
but also for guiding query mutation to explore a broad range of distributed behaviors,
as we will see next.

\subsection{SDM-Guided Query Mutation }\label{subsec:mutation}

With the SDM in place, the next step is to generate test SQL queries. 
We start from base queries that are either generated by existing tools~\cite{dynsql,SQLsmith} 
or observed to be executable during testing. 
For the former,
we initialize them to access only local tables on the same node, i.e., from the ``Local on the same node'' row of the SDM. 
Such queries are unlikely to be rejected because this setting closely resembles a centralized DBMS.
%(except that communication occurs only between that node and the coordinator).
 
% our next step is to generate test SQL  queries.
%We first construct a \zz{an original} base  query that accesses only local tables on the same node,
%i.e., tables from the ``Local on the same node''  row of the SDM.
%Such queries can be produced by existing  generators~\cite{dynsql,SQLsmith} and are unlikely to be rejected by the DDBMS under test, as this setting closely resembles a centralized DBMS
%(except that communication occurs only between that node and the coordinator).

%, which is equivalent to a centralized DBMS setting.
%Such queries can be produced by existing query generators~\cite{dynsql,SQLsmith} and are  unlikely to be rejected by the DDBMS under test due to the local setup.
%Note that, when executed, 
%these queries can already enable a special distributed execution in which query processing is localized to a single node, with communication only between that node and the coordinator.

Given a base query, \ourapproach proceeds by applying a series of mutations to it, aimed at triggering diverse distributed query executions and optimizations.
Underlying this process is a generic mutator
$\mu = (\rho, \Delta)$, 
where  $\rho$ maps each table instance  $\mathcal{M}[d, s]$  in the base query to another
(or the same)
table instance $\mathcal{M}[d', s']$  in the SDM, and
$\Delta$ denotes a set of additional (optional) transformations to the base query  tailored to specific distributed query optimizations.

\begin{figure}[t]
  \centering
  \includegraphics[width=\columnwidth]{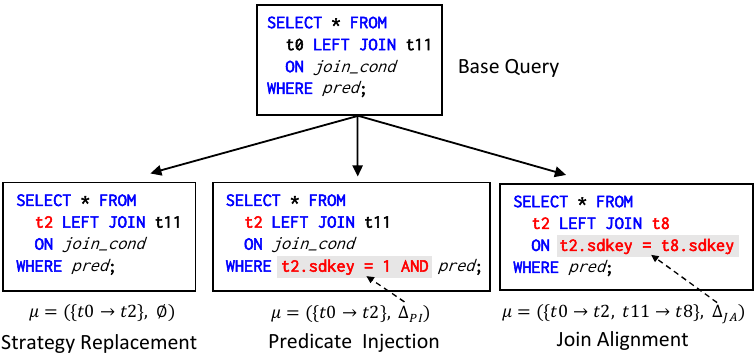}
  \captionsetup{skip=8pt}
  \caption{Examples of three types of query mutations, along with  corresponding $\rho$ and $\Delta$ specifications. }
  \label{fig:mutate}
\end{figure}

In the following,
we  present three instances of $\mu$,
with corresponding examples shown in 
Figure~\ref{fig:mutate}.
The first serves as a general mutation for exploring diverse distributed behaviors, while the remaining two target specific distributed query optimizations.
 We assume that the base query is 
executable in  the  DDBMS under test.

\inlsec{Strategy Replacement (SR)}
In this mutation, $\rho$ replaces tables in a query with other tables from the same column of the SDM (i.e., $s = s'$) but with different distribution strategies  (i.e., $d \neq d'$).
Since the new tables share the same schema as the original ones, the mutated query remains syntactically valid. 
Oftentimes, changing the distribution strategy of even a single table  can already alter  distributed query execution behavior.
For example, $\rho$ can be used to trigger different join optimizations across table types, e.g., between a reference table and a distributed table. 
Note that SR does not involve additional transformations  to the base query, i.e., $\Delta_\mathit{SR} = \emptyset$.

\begin{example}
In Figure~\ref{fig:mutate} (left branch), $\rho$ replaces \sqlvar{t0} with \sqlvar{t2}.
Before the mutation, \sqlvar{t0} and
(a copy of)
the reference table \sqlvar{t11} can be joined locally on the same node.
After the mutation, the join structure is preserved, but its execution semantics change, as \sqlvar{t0} is replaced by the
distributed table \sqlvar{t2} 
(recall from Section~\ref{subsec:ds} that co-located tables are a special case of distributed tables).
Consequently, the join may no longer be executed locally on the same node.
\end{example}

%Shard Routing. Scanning a distributed table often requires accessing all of its shards. However, when a query includes a predicate on the sharding key, execution can be restricted to the shards that satisfy the predicate. This optimization is known as shard routing (or shard pruning) [8, 50].

\inlsec{Predicate Injection (PI)}
This mutation is designed to trigger shard routing, a distributed query optimization that limits query execution to a subset of relevant shards
rather than
all shards.
In addition to introducing at least one distributed table (via $\rho$ guided by the SDM),
 this mutation applies
\(\Delta_{\mathit{PI}}\)
which inserts a predicate into the \sqlkw{WHERE} clause of the base query.
In particular, the predicate must be defined on the sharding key of the distributed table and inserted into the filter condition associated with that table.

\begin{example}
In Figure~\ref{fig:mutate} (middle branch), 
after $\rho$ replaces \sqlvar{t0} with \sqlvar{t2}, 
\(\Delta_{\mathit{PI}}\)
 inserts the predicate \sqlvar{t2.sdkey = 1} into the \sqlkw{WHERE} clause,
 making the sharding key of \sqlvar{t2} explicitly appear in the filter condition.
When processing this query, the optimizer can infer that only the shards containing rows with \sqlvar{t2.sdkey = 1} are relevant.

\end{example}

%A join between two distributed tables often requires cross-shard communication. However, if the tables are co-located, a join on the sharding key can be executed locally, as rows with the same sharding key are placed on the same node

\inlsec{Join Alignment (JA)} 
This mutation targets co-located joins, a  query optimization that executes joins locally between distributed tables without cross-shard communication.
\(\Delta_{\mathit{JA}}\)  applies only when all table instances participating in a \sqlkw{JOIN} are already co-located after  applying $\rho$ (based on the SDM).
It then rewrites the join conditions so that joins between the participating tables are explicitly expressed as equalities on their common sharding keys.

\begin{example}
In Figure~\ref{fig:mutate} (right branch), 
after applying $\rho$, \sqlvar{t0} and \sqlvar{t11} are replaced with \sqlvar{t2} and \sqlvar{t8}, respectively.
\(\Delta_{\mathit{JA}}\) 
 then rewrites the \sqlkw{ON} predicate as \sqlvar{t2.sdkey = t8.sdkey}, where \sqlvar{t2} and \sqlvar{t8} are co-located on the sharding key
 as determined during SDM construction.
Consequently, the join is transformed into a form that is likely to be executed as a co-located join.
\end{example}

\begin{figure}[t]
  \centering
  \includegraphics[width=\columnwidth]{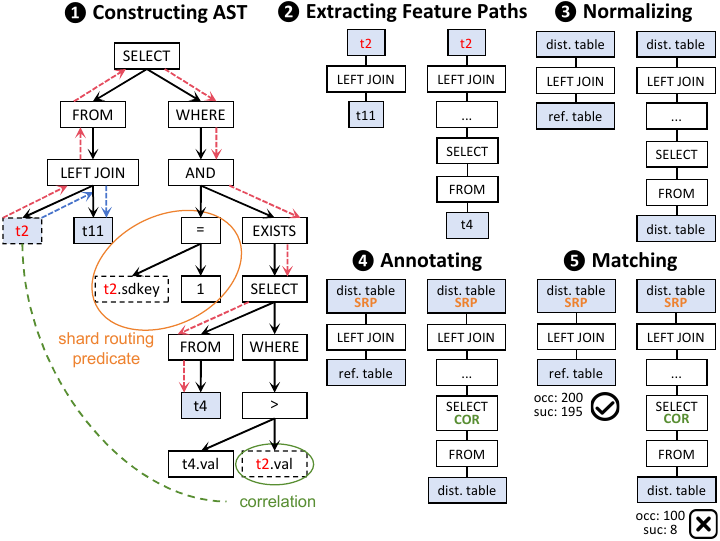}
  \captionsetup{skip=5pt}
  \caption{An illustration of the  history-guided feedback mechanism using the SQL query in Figure~\ref{fig:limitation}.
}
  \label{fig:feedback}
\end{figure}

\subsection{History-Guided Feedback}\label{sec:feedback}
While  the mutation diversifies generated queries, many may be rejected by the DDBMS due to unsupported SQL features (Section~\ref{sec:CapabilityBoundary}),  reducing overall testing efficacy.
To address this, we introduce a feedback mechanism that guides the generation toward executable queries
by leveraging historical successful and failed cases. 

Figure~\ref{fig:feedback} 
illustrates our  history-guided feedback mechanism using the SQL query in Figure~\ref{fig:limitation}.
%its pseudocode is provided in Algorithm~\ref{alg:feedback}. 
It begins by parsing the \emph{abstract syntax tree} (AST)~\cite{aho2006} of a mutated  query candidate,
which provides a structured representation of the query's syntactic constructs.
Formally,
the AST of a SQL query $Q$ is defined as 
$A(Q) = (V,E)$,
where $V$ is the set of AST nodes
representing syntactic constructs in $Q$,
including
statement constructs (e.g., \sqlkw{SELECT}),
table-reference constructs (e.g., \sqlvar{t2}),
relational operator constructs (e.g., \sqlkw{LEFT JOIN}),
and
expression constructs (e.g., the \sqlkw{AND} predicate);
$E  \subseteq V \times V$ 
is the set of directed edges capturing  parent–child relationships between nodes. 
The AST corresponding to the query in Figure~\ref{fig:limitation} is shown in Step~\ding{182},
which
 can be constructed using existing approaches~\cite{dynsql,PQS}. 

We next describe the remaining four core steps, 
which  determine whether a mutated query candidate 
qualifies as a test query.

\inlsec{\ding{183} Extracting  Feature Paths} 
We aim to avoid generating mutated queries that are similar to those previously rejected. 
A key challenge, therefore, is how to represent such similarity.

Our insight is that if a base query is executable but a mutated query is not, 
the cause of rejection likely stems from unsupported features introduced by the replaced tables,
as the mutation preserves the schema but alters their distribution strategies,
along with the associated transformations (as defined in $\mu$).
To capture this, we define \emph{feature paths}.
Intuitively, a feature path captures the structural relationship between two table nodes in the AST, encoding the sequence of relational operators and predicates that connect them, 
and thus characterizes how their distribution strategies jointly affect the executability of the query.

%\todo{and thereby how their distribution strategies are composed during execution.}
%\zz{Intuitively, a feature path captures the structural relationship between two table nodes in the AST, encoding the sequence of relational operators and predicates that connect them, and therefore represents a sequence of combined operations under the constraint of typical distribution strategies.}

Feature paths are extracted by starting from each mutated table node and 
traversing the AST, viewed as an undirected graph, to other table nodes.
All intermediate nodes along such paths correspond to SQL constructs.
%, except for the endpoints.
As the AST is a tree, the path between any two table nodes is unique.
Note that 
paths between unchanged tables  are not extracted, 
as they are assumed to remain executable given that they already appear in the original executable query.
As shown in Figure~\ref{fig:feedback} (\ding{183}),
   two feature paths are extracted
 from the AST by traversing the blue and red paths, respectively, both starting from the mutated table \texttt{t2}.

\inlsec{\ding{184} Normalization}
Queries that exhibit similar distributed execution behaviors may nonetheless produce different feature paths,
because the above representation retains concrete table identities.
For example, the execution of \sqlvar{t2} \sqlkw{LEFT JOIN} \sqlvar{t11} may be similar to that of 
\sqlvar{t9} \sqlkw{LEFT JOIN} \sqlvar{t5} in the DDBMS,
as the involved tables share the same distribution strategies  
(i.e., 
a distributed table joined with a reference table; see Figure~\ref{fig:sdm}),
yet are treated as distinct feature paths due to their different names. 
This introduces unnecessary variability,
which inflates the feature path space, 
leads to a lower matching rate with historical patterns in Step~\ding{186}, 
and
ultimately
weakens the effectiveness of the feedback guidance.

%This limitation arises because the original feature paths preserve concrete table references. 
%However, in DDBMSs, execution behavior is primarily determined by the distribution strategies of tables rather than their identities. 
%For example, the execution of \sqlvar{t2} \sqlkw{LEFT JOIN} \sqlvar{t11} may be similar to that of 
%\sqlvar{t9} \sqlkw{LEFT JOIN} \sqlvar{t11}, 
%as long as the involved tables have the same distribution roles 
%(e.g., a distributed table joined with a reference table), 
%even though the table names differ.

To address this issue, we normalize feature paths based on the 
\emph{effective distribution roles} of table references, 
rather than their concrete names. 
The effective role of a table reference is the distribution strategy it exhibits in the given query context, as determined by its SDM configuration and the surrounding SQL structure.
For example, \sqlvar{t2} behaves as a distributed table, 
as shown in  Figure~\ref{fig:feedback} (\ding{184}), 
despite originating from a co-located table in Figure~\ref{fig:sdm}, as it does not participate in any co-located join.

After normalization,
two feature paths are considered equivalent if they share the same sequence of non-table SQL constructs 
and their endpoint table references have the same effective distribution roles. 
This yields a canonical representation,
enabling matching across queries with similar distributed execution behaviors.

\inlsec{\ding{185} Annotation}
Another limitation of original feature paths is that they fail to capture execution-relevant information outside the path. 
For example, a shard-routing predicate (\sqlvar{t2.sdkey = 1}) may restrict access to \sqlvar{t2} to only a subset of its shards during distributed execution, rather than all shards. 
Such information may significantly affect execution behavior (e.g., by reducing cross-shard communication), 
but is not reflected in the feature path itself.

To address this issue, we augment feature paths with
\emph{node-level annotations} 
to capture execution semantics beyond structural information, 
thereby enabling more accurate matching. 
In our current design, we consider two annotation families most relevant to distributed execution behaviors in DDBMSs:  
\emph{optimization annotations} and \emph{correlation annotations}.

\paragraph{Optimization Annotations}
These  annotations capture distributed optimizations associated with tables or operators in a query, 
such as shard routing and co-located joins,
 which are frequently triggered by our mutations. 
For example,
for shard routing,
we record routing predicates and annotate the corresponding table node with ``SRP''. 
For example,
the starting node of each path in  Figure~\ref{fig:feedback}(\ding{185}) is annotated 
to indicate that a shard-routing predicate is applied to \sqlvar{t2}. 
Similarly, 
for co-located joins, normalizing participating tables into co-located tables is insufficient, 
as multiple \sqlkw{JOIN}s may appear along a path. 
Therefore, we  annotate the specific \sqlkw{JOIN} nodes that perform co-located joins
with ``CLJ''.

\paragraph{Correlation Annotations}
A correlation occurs when a subquery references a table in an outer query block. 
For example,  \sqlvar{t2}  is referenced in an \sqlkw{EXISTS} subquery (via the predicate \sqlvar{t4.val > t2.val}), creating a dependency between the outer query and the subquery. 
Such dependencies may introduce additional execution and communication overhead in DDBMSs, e.g., when subquery evaluation depends on values from the outer query; 
moreover, they
 may require complex distributed semi-join or anti-join execution across shards, 
 potentially causing an otherwise executable query to be rejected.
Accordingly, correlation annotations capture such cross-block dependencies by annotating the corresponding  nodes with ``COR'',
e.g., the \sqlkw{SELECT} node in  Figure~\ref{fig:feedback}(\ding{185}), 
so that correlated paths can be distinguished during matching.

\smallskip
Note that, following Step~\ding{184}, 
feature path equivalence is then redefined by incorporating these annotations.

\inlsec{\ding{186} Matching against Historical Paths}
\ourapproach maintains a path set $S_H$ that stores feature paths extracted from all previously executed  test queries, 
together with their execution statistics.
Each element in $S_H$ is represented as a tuple $(p, \mathit{occ}, \mathit{suc})$, 
where $p$ is a feature path obtained after normalization and annotation, 
$\mathit{occ}$ is its occurrence count, 
and $\mathit{suc}$ is its success count. 
Two paths are considered a \emph{match} if they are equivalent, 
independent of their associated statistics.

Let $Q$ be a mutated query candidate, and let $\mathcal{P}(Q)$ denote its set of
(normalized and annotated)
feature paths. 
For each $p \in \mathcal{P}(Q)$, 
if there exists an entry $s = (p', \mathit{occ}, \mathit{suc}) \in S_H$ 
such that $p$ matches $p'$,
\[
\mathit{occ}(s) \geq k
\quad\text{and}\quad
\frac{\mathit{suc}(s)}{\mathit{occ}(s)} \leq \theta
\]
then $Q$ is filtered out  by \ourapproach; 
otherwise, $Q$ is submitted to the DDBMS for execution.
Here, $k$ is the minimum 
occurrence threshold
and $\theta$ is the success-rate threshold. 
In practice, we choose a sufficiently large $k$ (e.g., 10) to ensure that decisions are based on reliable historical evidence. 
Moreover, some unsupported features may not always lead to query rejection, e.g., due to optimizer pruning. 
Therefore, a path is considered unsupported only when its empirical success rate is consistently below  $\theta$. 
In our experiments, we set $\theta = 10\%$, and the rationale is provided in Section~\ref{sss:theta}.

As shown in  Figure~\ref{fig:feedback}(\ding{186}), two feature paths are matched in the history. 
One has a high empirical success rate ($97.5\%$), 
 while the other has a very low success rate  ($8\%$)
due to the unsupported  feature shown in Figure~\ref{fig:limitation}. 
Therefore, this  candidate query is filtered out by \ourapproach prior to its execution in the DDBMS.

\begin{table*}[t]
 \captionsetup{skip=5pt}
  \caption{Summary of the tested DDBMSs.}
  \label{tab:DDBMSS}
  \centering
  %\tablesize
      \resizebox{\textwidth}{!}{
  \begin{tabular}{lccrrcc}
  %{p{1.6cm} p{1.4cm} p{2.0cm} p{0.9cm} p{0.7cm}}
    \toprule
    \textbf{DDBMS} & \textbf{Architecture} & \textbf{Base} & \textbf{GitHub} & \textbf{LoC} 
    % &\textbf{Languages} 
    &\textbf{Initial} 
    & \textbf{Tested Version} \\
 &  & \textbf{DBMS} & \textbf{Stars} & 
    % &\textbf{Languages} 
    &\textbf{Release} 
    &  \\
    \midrule
    Citus &
      extension &
      PostgreSQL &
      12.7K &
      350K
      & 2012 
      % & C++
      & v12.5 w/ PostgreSQL 16.2; v13.2 w/ PostgreSQL 17.2; v14.0 w/ PostgreSQL 18.2\\
    ShardingSphere &
      middleware&
      PostgreSQL %, MySQL
      &
      20.8K &
      790K
      & 2016 
      % & Java
      & v5.5.2 w/ PostgreSQL 17.0\\
    Vitess &
   cloud-native &
      MySQL &
      21.2K &
      210K
      & 2010 
      % & Go
      & v24.0.0 w/ MySQL 8.4.8\\
    ClickHouse &
      cloud-native&
      -- &
      49K &
      1.1M
      & 2012 
      % & C++ 
      & v25.12.4.35\\
    % Tidb &
    %   Cloud-native&
    %   / &
    %   39.7k &
    %   0.8M\\
    \bottomrule
  \end{tabular}
  }
\end{table*}

\section{Evaluation }
\label{sec:evaluation}
We implement our  
\ourapproach approach as a tool called \mytool. 
It builds on the codebase of DynSQL~\cite{dynsql}, 
which provides a basis for generating syntactically complex and valid SQL queries that can be  executed on centralized DBMSs,\footnote{These queries involve
common SQL constructs, e.g., DQL and DML, and advanced features, including subqueries, aggregation functions, and window functions.} 
as well as diverse database schemas.
The overall codebase consists of 17 KLOC in C++, including 1.6 KLOC for 
the query mutator and the feedback mechanism,
1.9 KLOC for DDBMS instance generation (i.e., SDMs) and bug checking, 
and 5.7 KLOC for supporting multiple DDBMSs. 
%While 
%\nobi{the mutation and guidance logic} 
%is reusable across DDBMSs, 
%adapting \mytool{} to a new system necessitates implementing \nobi{system-specific distribution strategy generators and strategy maintaining logic, which ensures that generated states and subsequent mutations remain distribution aware
%throughout the testing process.}
%\nobi{mention our focus on two opts and defer discussion}

%\zz{We realize the history path set as a \emph{prefix tree} as an implementation optimization.
%This reduces space overhead by sharing common prefixes across paths, and supports more efficient history maintenance by storing $occ$ and $suc$ along the tree structure during traversal.}

We conduct an extensive evaluation of \mytool{} and compare it against  state-of-the-art testers for DDBMS bug detection.
We aim to address the following questions:

%To understand the effectiveness of \mytool, we evaluate it on real-world and production-level DDBMSs. Our evaluation aims to answer the following questions:

\begin{description}[leftmargin=18pt]
    \item[Q1.] Can \mytool{} uncover previously unknown bugs in production DDBMSs (Section~\ref{sec:DiscoverBugs})?
    \item[Q2.] Does \mytool{} outperform the state-of-the-art in generating high-quality queries
    %triggering distributed query optimizations, 
    and 
    identifying bugs (Section~\ref{sec:Comparison})?
    \item[Q3.] How does each of our key design choices contribute to \mytool's effectiveness gains (Section~\ref{sec:Sensitivity})?
\item[Q4.] What are the resource overheads of \mytool{} (Section~\ref{sec:overhead})?
\end{description}

%\RQ{Can \mytool{} find bugs in real-world DDBMSs by distribution aware testing approach? (Section~\ref{sec:DiscoverBugs})}
%\RQ{Does \mytool{} perform better than other state-of-the-art testing tools on Testing DDBMSs? (Section~\ref{sec:Comparison})}
%\RQ{How does each component of our approach contribute to \mytool{} in testing DDBMSs? (Section~\ref{sec:Sensitivity})}
%\RQ{(Interpretability). Do the invalid dependency patterns identified by \mytool{} align with the DDBMS’s unsupported features? (Section~\ref{sec:interpret})

% \RQ{How effective is \mytool{} compared to other method in detecting the bugs?}
% \RQ{How effective is \mytool{} compared to other method in detecting the bugs?}

\subsection{Experimental Setup}
%\subsubsection{Tested DDBMSs.}
We evaluate recent releases of four widely used production DDBMSs of different kinds. 
Citus~\cite{citus}
extends PostgreSQL into a distributed database, enabling horizontal scaling across multiple nodes.
Apache ShardingSphere~\cite{shardingsphere}
operates as a database middleware layer (e.g., on top of PostgreSQL), aiming to maximize the computing capabilities of existing databases.
Vitess~\cite{vitess}
is a cloud-native distributed database system built around MySQL, supporting virtually unlimited scalability through generalized sharding.
ClickHouse~\cite{clickhouse}
is a standalone, columnar DBMS for online analytical processing (OLAP), enabling real-time analytical queries over large datasets.
Table~\ref{tab:DDBMSS} summarizes their key characteristics,
including  architectures, popularity, codebase sizes, and tested versions.

%We evaluate \mytool{} on widely-used and open-source DDBMSs of the latest versions for convenient bug reporting, as shown in table~\ref{tab:DDBMSS}.
%We tested two PostgreSQL-based DDBMSs, Citus~\cite{citus} and Apache ShardingSphere~\cite{shardingsphere}. Citus is an extension of PostgreSQL. Apache ShardingSphere is a distributed middleware that uses PostgreSQL or MySQL as the backend.
%We additionally tested ClickHouse~\cite{clickhouse} to evaluate generalizability.

% To evaluate generalizability, we additionally tested \mytool{} on two \emph{native} distributed databases, TiDB and ClickHouse. Both systems support a broad range of SQL constructs; consequently, even with arbitrary table layouts, query validity exceeds 90\%. Most failures are due to AST-level syntax errors rather than distributed query limitations, and queries are rarely rejected for being distributed-unoptimizable. In such systems, we expect our feedback to have limited impact. Instead, our goal is to stress how the \emph{same} query behaves under different  configurations and whether changes in distributed semantics expose bugs. Although TiDB and ClickHouse offer fewer user-configurable options than the other DDBMSs we studied, \mytool{} still found one bug in TiDB and two bugs in ClickHouse, all of which are tied to distributed execution logic.

We deploy each DDBMS cluster using both Docker Compose and Kubernetes. 
For Citus and ShardingSphere, each cluster consists of one coordinator node and five data nodes. 
For Vitess, we employ five shard nodes and two primary nodes. 
For ClickHouse, we deploy two database nodes and three keeper nodes.
For differential testing, 
we construct, for each distributed deployment, a corresponding centralized deployment on a single node.  
In particular, for Citus, ShardingSphere, and Vitess, the centralized deployment corresponds to their underlying base DBMSs (see Table~\ref{tab:DDBMSS}).

For each SDM, we randomly generate 4--9 schemas and 4--6  distribution strategies of different types.
Each generated table contains 40--120 rows chosen uniformly at random. 
We regenerate the SDM after every 1K generated queries, and each query is mutated for up to 10 times.
While larger parameter settings (e.g., more schemas or mutation times) may expose additional bugs, 
the above settings already suffice to uncover many bugs in practice, as we will see.
%While larger parameter settings (e.g., more schemas or mutation steps) may expose additional bugs, 
%they also incur higher overhead for \mytool. 
%Hence, there is no single optimal configuration. Nevertheless, our relatively small settings  suffice to uncover a large number of bugs in practice. 
%Finally, we set $\theta = 10\%$;
% the rationale is provided in Section~\ref{sss:theta}.

All experiments were conducted on a 64-bit Ubuntu 22.04 machine with an AMD Ryzen 5975WX (32 cores) and  256 GB RAM. 
Each data point is averaged over five runs.

\begin{table}[t]
  \captionsetup{skip=5pt}
  \caption{Summary of the bugs detected by \mytool, with distribution-related bugs shown in parentheses.}
  \label{tab:bug}
  %\tablesize
  \centering
  \small
   %\resizebox{\columnwidth}{!}{
  \begin{tabular}{l|ccc}
  %{p{2.0cm} C{1.4cm} C{1.4cm} C{1.4cm}}
    \toprule
    \textbf{DDBMS} & \textbf{\#detected} & \textbf{\#confirmed} & \textbf{\#fixed} \\
    \midrule
     Citus          & 8 (8)     & \textcolor{black}{7 (7)} & \textcolor{black}{7 (7)} \\
     ShardingSphere & 9 (7)     & \textcolor{black}{8 (6)} & \textcolor{black}{7 (5)} \\
     Vitess         & 12 (12)   & \textcolor{black}{5 (5)} & \textcolor{black}{1 (1)} \\
     ClickHouse     & 2 (1)     & \textcolor{black}{1 (0)} & 0 (0)  \\
     \midrule
     Total          & 31 (28)   & \textcolor{black}{21 (18)} & \textcolor{black}{15 (13)}  \\
    \bottomrule
  \end{tabular}
  %}
\end{table}

% \begin{figure}[t]
%   \centering
%   \includegraphics[width=\columnwidth]{figures/bug_count_clustered_stacked.png}
%   \caption{Bugs Detected by \mytool.}
%   \label{fig:buglist}
% \end{figure}

\subsection{Uncovering New Bugs}\label{sec:DiscoverBugs}
Overall, we reported 31 previously unknown bugs across the four DDBMSs, 
as summarized in Table~\ref{tab:bug}. 
A complete list of these bugs, including the corresponding DDBMS, detailed description, bug type, status, and involved SQL features, is provided in the Appendix~\cite{tech-rpt}.
%A list of these bugs, along with detailed descriptions, is provided in the Appendix~\cite{tech-rpt}. %Appendix~\ref{app:bugs}. 
Among them, 28 
%are distribution-specific, whose  manifestations  or root causes 
are related to 
distribution-specific mechanisms across the query processing pipeline, including rewriting, planning, optimization, and execution,
as well as distribution strategies.
This demonstrates \mytool's effectiveness in  
triggering distribution-specific bugs beyond those  observed
in centralized DBMSs. 
The remaining three bugs include one timestamp format inconsistency %in ShardingSphere 
 and two SQL parser issues.
%Note that some of these bugs have long persisted, 
%e.g.,  persisted for nearly three years
%across six major versions 
%despite extensive prior testing efforts
%\nobi{since the initial releases of the tested DDBMSs,}
%before being uncovered by \mytool.
%Note that some of these bugs persisted for a long time before being uncovered by \mytool, 
%e.g.,
%the one shown in Figure~\ref{fig:intro-bug} remained undetected for three years across six major versions. 
Notably, some of these bugs remained undetected for years,
e.g.,
half of the bugs in Citus went undetected for
over four years.
We present two additional representative bugs in the Appendix~\cite{tech-rpt},
complementing the one shown in Section~\ref{intro}.

%As of this writing, 
%\nobi{29} bugs have been confirmed by the developers, with 10 already fixed in the latest releases. 
%All bugs reported for Citus, ShardingSphere, and Vitess have been confirmed. 
%Due to longer response times from the ClickHouse developers, none of the reported  bugs there
% have been confirmed yet; 
% however, confirmations (of all the bugs) from the other three systems increase our confidence that these bugs are also genuine.

As of this writing,
21 bugs have   been confirmed by the developers, 
with 15 already fixed in recent releases.
%Due to slower responses from the Vitess and ClickHouse teams, 
%many bugs are still under investigation and have yet to be confirmed.
%Nevertheless, 
%The confirmed cases increase our confidence that 
%the remaining ones are also genuine.
% Moreover, 
Our test oracles are highly unlikely to produce false positives, as they rely on clear signals such as crashes, timeouts, errors,  unexpected exceptions, and  discrepancies in the results of the same query under centralized and distributed executions of the \emph{same} system. 
%Table~\ref{tab:bug-types} shows the distribution of the detected bugs by type: 
%half are logic bugs causing incorrect query results, which are
%uncovered via differential testing;
%the rest are identified via log inspection and exception monitoring.
We provide the
distribution of detected bugs by type
in the Appendix~\cite{tech-rpt},
where half of them are logic bugs.

\begin{table*}[t]
  \begin{minipage}[t]{0.57\textwidth}
    %\vspace{0pt}
    \captionsetup{skip=10pt}
    \caption{Bugs detected over 24 hours of testing. SQLancer's current implementation only supports testing Citus and ClickHouse.}
    \label{tab:24hourBug-compare}
    \centering
    \scriptsize
    \resizebox{\linewidth}{!}{%
      \begin{tabular}{l|cc|cccc|cccc}
        \toprule
        \textbf{DDBMS} & 
        \multicolumn{2}{c}{\textbf{\mytool}} &
        \multicolumn{2}{|c}{\textbf{SQLancer}} &
        \multicolumn{2}{c}{\textbf{EET}} &
        \multicolumn{2}{|c}{\textbf{\mytool$_\mathit{!MT}$}} &
        \multicolumn{2}{c}{\textbf{\mytool$_\mathit{!FB}$}}
        \\
        \cmidrule(lr){2-3}\cmidrule(lr){4-5}\cmidrule(lr){6-7}\cmidrule(lr){8-9}\cmidrule(lr){10-11}
        & \textbf{total} & \textbf{dist.}
        & \textbf{total} & \textbf{dist.}
        & \textbf{total} & \textbf{dist.}
        & \textbf{total} & \textbf{dist.} 
        & \textbf{total} & \textbf{dist.} 
        \\
        \midrule
        Citus &
        3 & 3 &
        0 & 0 &
        0 & 0 &
        3 & 3 &
        1 & 1 
        \\
        ClickHouse &
        2 & 1 &
        0 & 0 &
        1 & 1 &
        2 & 1 &
        2 & 1 
        \\
        ShardingSphere &
        4 & 2 &
        -- & -- &
        2 & 1 &
        2 & 0 &
        3 & 1 
        \\
        Vitess &
        7 & 7 &
        -- & -- &
        2 & 2 &
        6 & 6 &
        5 & 5 
        \\
        \midrule
        Total &
        16  & 13 &
        0 & 0 &
        6 & 5 &
        13 & 10 &
        11 &  8
        \\
        \bottomrule
      \end{tabular}%
    }
  \end{minipage}%
  \hfill
  \begin{minipage}[t]{0.4\textwidth}
    \vspace{0pt}
    \centering

\begin{tikzpicture}
\begin{axis}[
    width=\textwidth,
    height=0.4\textwidth,
    xmin=0, xmax=24,
    ymin=0, ymax=17,
    xtick={1,4,8,12,16,20,24},
    ytick={0,4,8,12,16},
    xlabel={\footnotesize Time (hours)},
    ylabel={\footnotesize Cumulative  bugs},
    tick label style={font=\footnotesize},
    label style={font=\footnotesize},
    ylabel style={font=\footnotesize,yshift=-0.2em},
    xlabel style={font=\footnotesize,yshift=0.4em},
    tick style={draw=none},
    legend image code/.code={
    \draw[mark repeat=2,mark phase=2,#1]
    plot coordinates {
        (0cm,0cm)
        (0.18cm,0cm)
        (0.36cm,0cm)
    };
    },
    legend style={
        font=\scriptsize,
        draw=none,
        fill=none,
        at={(0.5,1.0)},
        anchor=south,
        cells={anchor=west},
        row sep=1pt,
    },
    legend columns=3,
    axis line style={black},
    tick align=outside,
    xtick pos=bottom,
    ytick pos=left,
    grid=none,
    clip=false,
]

% Draw baselines first.
% \addplot+[thick, brown, mark=diamond*, mark size=1.4pt, dashdotted,
%     mark options={fill=brown, draw=brown}]
%     table[x=time, y=NoMT, col sep=comma]
%     {figures/cumulative_bug_plot/cumulative_bug_data.csv};
% \addlegendentry{\mytool{}$_{\mathit{!MT}}$}

% \addplot+[thick, black, mark=otimes*, mark size=1.2pt, dotted,
%     mark options={fill=black, draw=black}]
%     table[x=time, y=NoFB, col sep=comma]
%     {figures/cumulative_bug_plot/cumulative_bug_data.csv};
% \addlegendentry{\mytool{}$_{\mathit{!FB}}$}

\addplot+[thick, brown, mark=square, mark size=1.5pt,
    mark options={fill=brown, draw=brown}]
    table[x=time, y=SQLancer, col sep=comma]
    {figures/cumulative_bug_plot/cumulative_bug_data.csv};
\addlegendentry{SQLancer}

\addplot+[thick, blue, mark=triangle, mark size=1.5pt,
    mark options={fill=blue, draw=blue}]
    table[x=time, y=EET, col sep=comma]
    {figures/cumulative_bug_plot/cumulative_bug_data.csv};
\addlegendentry{EET}

% Draw DistRanger last so it stays on top.
\addplot+[thick, red, mark=o, mark size=1.5pt,
    mark options={fill=red, draw=red}]
    table[x=time, y=Distranger, col sep=comma]
    {figures/cumulative_bug_plot/cumulative_bug_data.csv};
\addlegendentry{\mytool{}}

% Right-side endpoint labels, formatted like colored axis ticks.
% Format: total (distribution-specific).
\draw[red, line width=0.45pt] (axis cs:24,16) -- (axis cs:24.18,16);
\node[font=\scriptsize, text=red, anchor=west] at (axis cs:24.28,16) {16 (13)};

% \draw[brown, line width=0.45pt] (axis cs:24,13) -- (axis cs:24.18,13);
% \node[font=\scriptsize, text=brown, anchor=west] at (axis cs:24.28,13) {13 (10)};

% \draw[black, line width=0.45pt] (axis cs:24,11) -- (axis cs:24.18,11);
% \node[font=\scriptsize, text=black, anchor=west] at (axis cs:24.28,11) {11 (8)};

\draw[blue, line width=0.45pt] (axis cs:24,6) -- (axis cs:24.18,6);
\node[font=\scriptsize, text=blue, anchor=west] at (axis cs:24.28,6) {6 (5)};

\draw[brown, line width=0.45pt] (axis cs:24,0) -- (axis cs:24.18,0);
\node[font=\scriptsize, text=brown, anchor=west] at (axis cs:24.28,0) {0 (0)};

\end{axis}
\end{tikzpicture}
\captionsetup{skip=0pt}
    
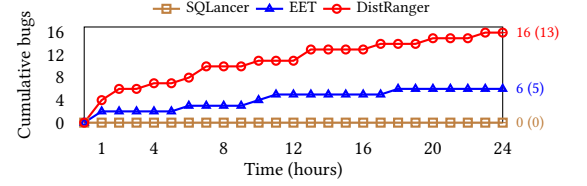
\captionof{figure}{Cumulative  bugs  detected in the four DDBMSs
over 24 hours. Distribution-related bugs are shown in parentheses.}
    \label{fig:cumulative_bugs2}
  \end{minipage}
\end{table*}

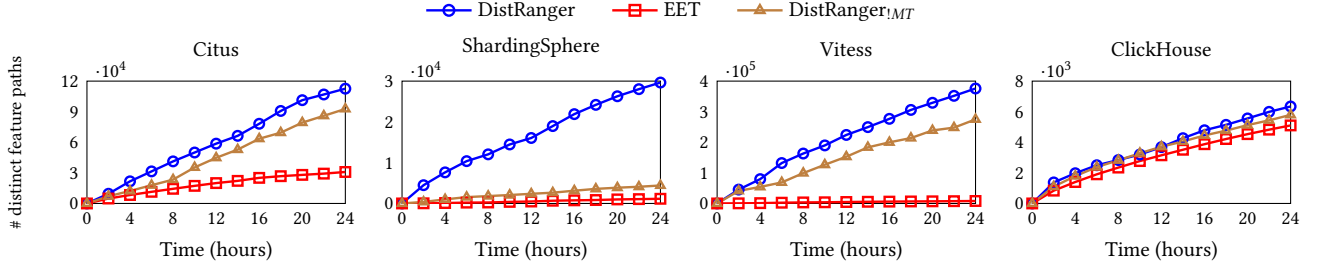
\begin{figure*}
  \centering
  \begin{tikzpicture}
    \begin{groupplot}[
      group style={
        group size=4 by 1,
        horizontal sep=0.75cm,
      },
      width=5cm,
      height=3.2cm,
      xmin=0, xmax=24,
      ymin=0, ymax=70000,
      xtick={0,10,20},
      ytick={0,10000,20000,30000,40000,50000,60000},
      xlabel={Time (hours)},
      xlabel style={font=\small},
      title style={font=\small},
      tick label style={font=\footnotesize},
      major grid style={dashed,gray!30},
      minor tick num=1,
      every axis plot/.append style={line width=0.9pt, mark size=1.8pt},
      tick style={draw=none},
      xtick={0,4,8,12,16,20,24},
    ]

      % -------- Citus --------
      \nextgroupplot[
          title={Citus},
          ymin=0, ymax=120000,
          ytick={0,30000,60000,90000,120000},
           scaled y ticks=base 10:-4,
        ]
      \addplot+[mark=o, blue] coordinates {(0,0) 
(2,9497) (4,21699) (6,31454) (8,41265) (10,49970) (12,58661) (14,66334) (16,78156) (18,90736) (20,101311) (22,106889) (24,112586)
      };
      \addplot+[mark=square, red] coordinates {(0,0) 
(2,4843) (4,8297) (6,11416) (8,14362) (10,17261) (12,19979) (14,22159) (16,25103) (18,26756) (20,28034) (22,29267) (24,30876)

      };
      \addplot+[mark=triangle, brown] coordinates {(0,0) 
(2,7032) (4,12269) (6,17876) (8,23384) (10,35335) (12,44776) (14,52832) (16,63476) (18,69437) (20,79289) (22,85973) (24,92698)
      };

      % -------- ShardingSphere --------
      \nextgroupplot[
          title={ShardingSphere},
          ymin=0, ymax=30000,
          ytick={0, 10000,  20000, 30000},
        ]
      \addplot+[mark=o, blue] coordinates {
        (0,0) (2,4481) (4,7622) (6,10377) (8,12042) (10,14467) (12,16054) (14,18974) (16,21903) (18,24181) (20,26287) (22,28028) (24,29651)
      };
      \addplot+[mark=square, red] coordinates {
        (0,0) (2,40) (4,105) (6,207) (8,236) (10,387) (12,475) (14,668) (16,764) (18,839) (20,989) (22,1064) (24,1145)
      };
      \addplot+[mark=triangle, brown] coordinates {
        (0,0) (2,374) (4,997) (6,1555) (8,1808) (10,2068) (12,2369) (14,2644) (16,3125) (18,3575) (20,3865) (22,4092) (24,4406)
      };

      % -------- Vitess --------
      \nextgroupplot[
        title={Vitess},
        ymin=0, ymax=400000,
        ytick={0,100000,200000,300000,400000},
      ]
      \addplot+[mark=o, blue] coordinates {
        (0,0) (2,45352) (4,79561) (6,132127) (8,163226) (10,189765) (12,224031) (14,249569) (16,276661) (18,305807) (20,329147) (22,351870) (24,375841)
      };
      \addplot+[mark=square, red] coordinates {
        (0,0) (2,383) (4,1065) (6,2478) (8,3396) (10,3875) (12,4763) (14,5479) (16,6022) (18,6547) (20,6995) (22,7489) (24,8075)

      };
      \addplot+[mark=triangle, brown] coordinates {
        (0,0) (2,41048) (4,54103) (6,68763) (8,99253) (10,126548) (12,152764) (14,183475) (16,199846) (18,213879) (20,238766) (22,247895) (24,274893)
      };

      % -------- ClickHouse --------
      \nextgroupplot[
        title={ClickHouse},
        ymin=0, ymax=8000,
        ytick={0,2000,4000,6000,8000},
        scaled y ticks=base 10:-3,
      ]
      \addplot+[mark=o, blue] coordinates {
        (0,0) (2,1375) (4,1976) (6,2499) (8,2839) (10,3201) (12,3683) (14,4268) (16,4787) (18,5140) (20,5560) (22,5990) (24,6342)
      };
      \addplot+[mark=square, red] coordinates {
        (0,0) (2,838) (4,1402) (6,1906) (8,2347) (10,2773) (12,3149) (14,3512) (16,3875) (18,4213) (20,4533) (22,4826) (24,5103)
      };
      \addplot+[mark=triangle, brown] coordinates {
        (0,0) (2,1103) (4,1790) (6,2379) (8,2823) (10,3286) (12,3711) (14,4052) (16,4450) (18,4763) (20,5122) (22,5436) (24,5792)
      };

    \end{groupplot}

    % ----- manual legend -----
    \coordinate (legendpos) at
      ($(group c1r1.north west)!0.7!(group c4r1.north east)+(0,12mm)$);

    \begin{axis}[
      at={(legendpos)},
      anchor=south,
      hide axis,
      xmin=0, xmax=1,
      ymin=0, ymax=1,
      width=0pt,
      height=0pt,
      scale only axis,
      legend columns=3,
      legend style={
        draw=none,
        font=\small,
        /tikz/every even column/.append style={column sep=0.4cm},
      },
    ]
      \addlegendimage{blue, mark=o, line width=0.9pt}
      \addlegendentry{\mytool{}}
      \addlegendimage{red, mark=square, line width=0.9pt}
      \addlegendentry{EET}
      \addlegendimage{brown, mark=triangle, line width=0.9pt}
      \addlegendentry{\mytool$_{\mathit{!MT}}$}
    \end{axis}

    % ----- shared y label -----
    \node[rotate=90, anchor=south] at ([xshift=-7mm]group c1r1.west)
      {\footnotesize  \# distinct feature paths};

  \end{tikzpicture}

  \captionsetup{skip=2pt}
  \caption{Comparison of query diversity, measured as the cumulative number of distinct feature paths extracted from the executable queries  on each DDBMS over 24 hours.}
  \label{fig:diversity}
\end{figure*}

\subsection{Comparison with State-of-the-Art}
\label{sec:Comparison}
We compare \mytool{} against two representative state-of-the-art DBMS testing tools, namely SQLancer~\cite{SQLancer,PQS} and EET~\cite{EET}.

We select SQLancer\footnote{Specifically, we use the latest version of SQLancer as of April 27, 2026~\cite{sqlancer_commit_d38ddcb}.} as a strong baseline for \emph{end-to-end} database testing
due to its  success in finding
numerous
bugs in production DBMSs over the years.
Its current implementation supports testing Citus and ClickHouse, 
 for which Ternary Logic Partitioning (TLP)~\cite{TLP} is used as the test oracle. 
TLP partitions a query into three derived queries whose results are combined and compared with the original query result; any mismatch indicates a bug.

EET is a recent testing tool that supports diverse queries by incorporating rich SQL features (e.g., subqueries and various join types).
%It relies on a metamorphic oracle for bug detection
Given a query, EET rewrites selected expressions into semantically equivalent but structurally more complex forms, whose results are expected to be identical.
Since EET also builds on DynSQL, its query generator is largely comparable to that of \mytool, enabling a fair comparison of
query executability  %(Section~\ref{sss:validity}) 
and  diversity %(Section~\ref{sss:diversity}) 
in DDBMSs. 
Moreover, the shared codebase allows us to readily equip EET with our SDM to generate DDBMS instances,
which %it does not natively support.
%This 
enables  comparisons between \mytool{} and  EET  across all target DDBMSs.
 
\subsubsection{Bug Detection} 
To examine how effective each tool is in bug detection, we conducted a 24-hour empirical analysis.
The results are shown in Table~\ref{tab:24hourBug-compare}. 
Overall, \mytool{} identifies a total of 16 bugs, including 13 related to
distributed query processing and optimization
 in DDBMSs,\footnote{Note  that Table~\ref{tab:bug}  reports a larger number of bugs, as our full testing campaign spanned several months.} significantly outperforming SQLancer and EET.
Notably, all bugs detected by EET are also found by \mytool.

%processing bugs in  DDBMSs that are related to distribution strategies and distributed  optimizations.

In addition,
we analyze how the bugs detected by each tool accumulates over time,
as shown in Figure~\ref{fig:cumulative_bugs2}. Compared with SQLancer and EET, \mytool{} provides two key advantages:
it discovers the largest number of bugs overall and continues to uncover new bugs throughout the testing period.
This demonstrates that \mytool{} can detect bugs more effectively over time.

These performance gaps arise from fundamental design differences.
Both SQLancer and EET primarily target centralized settings, which limits the quality of their generated queries in terms of (i) executability, preventing a large fraction of them from being executed in DDBMSs,
and (ii) diversity,
hindering their ability to trigger diverse  system behaviors.
A detailed analysis is provided below.

\begin{table}[t]
  \captionsetup{skip=5pt}
  \caption{Comparison of query executability rates over 24 hrs.}
  \label{tab:validity-compare}
  \centering
  %\small
%\tablesize
\resizebox{\columnwidth}{!}{
  \begin{tabular}{l|c|cc|c}
  %{p{1.6cm} | C{1.4cm} | C{1.2cm} C{1.2cm} C{1.3cm}  | C{2cm} C{2cm}}
    \toprule
    \textbf{DDBMS} & \textbf{\mytool} & \textbf{SQLancer}
    & \textbf{EET}
    % & \textbf{\mytool$_{\mathit{QPG}}$} & \textbf{\(\mytool_{\mathit{!MT}}\)} 
    & \textbf{\mytool$_\mathit{!FB}$} \\
    \midrule
     Citus & 87.6\% & 38.9\% & 25.3\%
     %& 70.1\% & 70.7\% 
     & 47.0\% \\
     ClickHouse & 88.7\% & 33.0\%  & 85.1\%
     %& 79.3\% & 79.8\% 
     & 84.5\% \\
     ShardingSphere & 72.0\% & -- & 3.1\%
     %& 43.6\% & 36.2\% 
     &  22.6\% \\
     Vitess  & 84.6\% & -- & 13.8\%
     %& 71.7\% & 73.2\% 
     & 55.3\% \\
     % Tidb & 97.9\% & 83.2\% & 97.6\% & 97.6\% \\
    \bottomrule
  \end{tabular}
  }
\end{table}

\subsubsection{Test Query Executability} 
\label{sss:validity}
Executability is crucial, as only executable queries can be used to exercise system behaviors and expose potential bugs.
We measure the executability rate of queries generated by the three tools across the four  DDBMSs.
As shown in Table~\ref{tab:validity-compare}, 
 \mytool{} significantly outperforms EET
 (by up to 22.2$\times$), as well as SQLancer (by up to 1.7$\times$ on the two systems it supports).
 This demonstrates the effectiveness of our feedback mechanism in guiding the generator toward queries accepted by DDBMSs. %, achieving approximately 3$\times$ higher  executability rates. 
%Since \mytool{} builds on the DynSQL query generator (Section~\ref{sec:impl}), which already produces high-quality queries without syntactic errors, we further exclude SQLancer-generated queries that fail due to syntax and type-conversion issues. 
%Even after this adjustment, SQLancer only achieves a validity rate of 50.17\% for Citus and 
%\todo{xx\%}
%for ClickHouse.
%This demonstrates the effectiveness of our feedback mechanism in guiding the query generator to produce queries that are accepted by DDBMSs.

Note that  \mytool's  executability rate varies across DDBMSs.
This is primarily due to differences in unsupported features across systems, as well as the limited time (24 hours) available for our feedback mechanism to learn and guide generation.
Moreover, achieving a 100\% executability rate is inherently challenging.
One reason is that queries containing unsupported features may still be accepted by the DDBMS due to optimizer branch pruning,
which can mask invalid substructures.
We defer a detailed analysis to Section~\ref{sss:theta}.

\subsubsection{Test Query Diversity}
\label{sss:diversity}
The diversity of generated queries is another key factor for effective DDBMS testing. 
A tester that repeatedly generates structurally similar queries may exercise only a limited portion of the distributed query processing logic, even if they are executable. 
Therefore, we evaluate whether \mytool{} can continuously explore diverse  queries over time, compared to EET
(as both tools share a common base query generator).

As described in Section~\ref{sec:feedback}, a feature path summarizes table-to-table structural relationships together with their effective distribution strategies. 
Accordingly, the number of unique feature paths serves as a proxy for the diversity of distributed execution patterns covered by a  tool. 
We measure this using the cumulative number of distinct feature paths extracted from executable queries generated during 24 hours on each DDBMS.
As shown in Figure~\ref{fig:diversity}, \mytool{} consistently covers more unique feature paths than EET across all the  DDBMSs. 
This demonstrates that \mytool{} enables broader exploration of distributed query execution behaviors.

\begin{comment}

\bigskip
Test query diversity is another important factor for effective DDBMS testing.
A tester that repeatedly generates structurally similar queries may exercise only a limited portion of the distributed query-processing logic, even if these queries are executable.
Therefore, we further evaluate whether \mytool{} can continuously explore diverse distributed query structures over time.

As described in Section~\ref{sec:feedback}, a feature path summarizes a table-to-table structural relationship together with their effective distribution strategies.
Thus, the number of unique feature paths serves as a metric for the diversity of distributed execution patterns covered by a testing tool.
We use the cumulative number of unique feature paths extracted from generated queries. 

For each tool, we run 24-hour testing campaigns on each DDBMS and record the cumulative number of distinct feature paths observed over time. As shown in Figure~\ref{fig:diversity}, \mytool{} consistently covers more unique feature paths than EET on all four DDBMSs.
This indicates that \mytool{} not only improves executability, but also helps the tester explore a broader space of distributed query executions.
    
\end{comment}

\subsubsection{Triggering Specific  Query Optimizations} \label{optimization_expriments}
Another  advantage of \ourapproach is its generality:
it can be instantiated to target specific distributed query optimizations, such as shard routing and co-located joins (Section~\ref{subsec:mutation}).
Accordingly, we compare how frequently three tools trigger these two optimizations. 
For each tool, we generate 10K executable queries per run across five runs under randomly generated DDBMS instances.
We instrument key entry points in the optimization logic to record when each optimization is triggered (see the Appendix~\cite{tech-rpt} for details); 
thus, trigger counts approximate its execution frequency. 
%For ClickHouse, we evaluate only shard routing, as it does not support co-located joins.
Since SQLancer does not support generating distributed tables for ClickHouse and thus cannot trigger the shard-routing logic, 
we randomly distribute the generated tables across the ClickHouse cluster to enable a meaningful comparison.

%As shown in Figure~\ref{fig:trigger-rate-both} (left and middle),
%\mytool{} triggers the two distributed query optimizations  more frequently than both SQLancer and EET,
%achieving at least a 2.44$\times$ increase (for co-located joins in Citus) 
%and up to a 538$\times$ increase (for shard routing in ShardingSphere).
%This is expected, since our query mutation is specifically designed to trigger these distributed optimizations.

\begin{figure*}[t]
  \centering

  \begin{minipage}[t]{0.2\linewidth}
    \vspace{0pt}
    \centering
    \begin{scaletikzpicturetowidth}{\linewidth}
     \begin{tikzpicture}[scale=\tikzscale]
      \begin{groupplot}[
        group style={
          group size=1 by 2,
          vertical sep=0cm,
          xticklabels at=edge bottom,
        },
        width=5.4cm,
        height=2.7cm,
        xmin=-0.5,
        xmax=2.5,
        xtick=data,
        xticklabels={Citus$_\mathit{SR}$ ,Citus$_\mathit{CJ}$ ,ClickHouse$_\mathit{SR}$ },
        ybar=1pt,
        ylabel near ticks,
        /pgf/bar width=5pt,
        enlarge x limits=0.10,
        legend style={
          at={(-0.2 ,1.42)},
          anchor=west,
          legend columns=2,
          draw=none,
          column sep = 3pt
        }
      ]
        \nextgroupplot[
        font=\small,
        ymin=10,
        ymax=500,
        ytick={10,250,500},
        axis y discontinuity=parallel,
        axis x line=bottom,
        x axis line style={dotted,-},
        axis y line=left,
        ylabel={Mult. factor ($\times$)},
        ylabel style={xshift=-0.45cm},
        % ymode=log,
        % log basis y={10},
        % log ticks with fixed point,
        % scaled y ticks=false,
      ]
        \addplot[draw=blue,
        nodes near coords={\pgfmathprintnumber[fixed,precision=2]{\pgfplotspointmeta}},
        every node near coord/.append style={font=\small, yshift=0.5pt}
        ] coordinates {
        (0,1.0)
        (1,1.0)
        (2,1.0)
      };
      \addplot[
        draw=red,
        fill=red,
        point meta=y,
        nodes near coords={\pgfmathprintnumber[fixed,precision=2]{\pgfplotspointmeta}},
        every node near coord/.append style={font=\small, yshift=0.5pt}
      ] coordinates {
        (0,7.05)
        (1,2.49)
        (2,359.5)
      };
      \legend{ \large SQLancer, \large \mytool{}}
      
      \nextgroupplot[
        font=\small,
        ymin=0,
        ymax=8,
        ytick={0,2,4,6},
        axis x line=bottom,
        axis y line=left,
        y axis line style={-},
        x tick label style={rotate=30, anchor=north east},
      ]
      \addplot[draw=blue,
        nodes near coords={\pgfmathprintnumber[fixed,precision=2]{\pgfplotspointmeta}},
        every node near coord/.append style={font=\small, yshift=0.5pt}
        ] coordinates {
        (0,1.0)
        (1,1.0)
        (2,1.0)
      };
      \addplot[
        draw=red,
        fill=red,
        point meta=y,
        nodes near coords={\pgfmathprintnumber[fixed,precision=2]{\pgfplotspointmeta}},
        every node near coord/.append style={font=\small, yshift=5.5pt}
      ] coordinates {
        (0,7.05)
        (1,2.49)
        (2,359.5)
      };

    \end{groupplot}
  \end{tikzpicture}
    \end{scaletikzpicturetowidth}
  \end{minipage}
  \hspace{5ex}
  %\hfill
  % \hspace{-0.06\columnwidth}
  \begin{minipage}[t]{0.3\linewidth}
    \vspace{0pt}
    \centering
    \begin{scaletikzpicturetowidth}{\linewidth}
      \begin{tikzpicture}[scale=\tikzscale]
      \begin{groupplot}[
        group style={
          group size=1 by 2,
          vertical sep=0cm,
          xticklabels at=edge bottom,
        },
        width=8cm,
        height=2.7cm,
        xmin=-0.8,
        xmax=6.8,
        xtick=data,
        xticklabels={Citus$_\mathit{SR}$,
        Citus$_\mathit{CJ}$,
        ShardingSphere$_\mathit{SR}$,
        ShardingSphere$_\mathit{CJ}$,
        Vitess$_\mathit{SR}$,
        Vitess$_\mathit{CJ}$,
        ClickHouse$_\mathit{SR}$
        },
        ybar=1pt,
        ylabel near ticks,
        /pgf/bar width=5pt,
        enlarge x limits=0.10,
        legend style={
          at={(0.5,1.25)},
          anchor=south,
          legend columns=2,
          draw=none,
          column sep=4pt
        },
      ]

      \nextgroupplot[
        font=\small,
        ymin=5,
        ymax=450,
        ytick={10,300,450},
        axis y discontinuity=parallel,
        axis x line=bottom,
        x axis line style={dotted,-},
        axis y line=left,
                ylabel={Mult. factor ($\times$)},
        ylabel style={xshift=-0.55cm},
        %ylabel={\Large Mult. Factor ($\times$)},
        %ylabel style={xshift=-0.55cm},
        % ymode=log,
        % log basis y={10},
        % log ticks with fixed point,
        % scaled y ticks=false,
      ]
      \addplot[draw=blue,
        nodes near coords={\pgfmathprintnumber[fixed,precision=2]{\pgfplotspointmeta}},
        every node near coord/.append style={font=\small, yshift=0.5pt}
      ] coordinates {
        (0,1.0)
        (1,1.0)
        (2,1.0)
        (3,1.0)
        (4,1.0)
        (5,1.0)
        (6,1.0)
      };
      \addplot[draw=red, fill=red,
        nodes near coords={\pgfmathprintnumber[fixed,precision=2]{\pgfplotspointmeta}},
        every node near coord/.append style={font=\small, yshift=0.5pt}
      ] coordinates {
        (0,90.26)
        (1,1.76)
        (2,355.55)
        (3,397.67)
        (4,45.70)
        (5,204.80)
        (6,4.79)
      };
      \legend{\large EET,\large \mytool}
\hspace{2ex}
      \nextgroupplot[
        font=\small,
        ymin=0,
        ymax=5,
        ytick={0,2,4},
        axis x line=bottom,
        axis y line=left,
        y axis line style={-},
        x tick label style={rotate=30, anchor=north east},
      ]
      \addplot[draw=blue,
        nodes near coords={\pgfmathprintnumber[fixed,precision=2]{\pgfplotspointmeta}},
        every node near coord/.append style={font=\small, yshift=0.5pt}
      ] coordinates {
        (0,1.0)
        (1,1.0)
        (2,1.0)
        (3,1.0)
        (4,1.0)
        (5,1.0)
        (6,1.0)
      };
      \addplot[
        draw=red,
        fill=red,
        point meta=y,
        nodes near coords={\pgfmathprintnumber[fixed,precision=2]{\pgfplotspointmeta}},
        every node near coord/.append style={font=\small, yshift=2pt}
      ] coordinates {
        (0,90.26)
        (1,1.76)
        (2,355.55)
        (3,397.67)
        (4,45.70)
        (5,204.80)
        (6,4.79)
      };

    \end{groupplot}
  \end{tikzpicture}
    \end{scaletikzpicturetowidth}
  \end{minipage}
% \hspace{-0.06\columnwidth}
  \hspace{6ex}
  \begin{minipage}[t]{0.3\linewidth}
    \vspace{0pt}
    \centering
    \begin{scaletikzpicturetowidth}{\linewidth}
      \begin{tikzpicture}[scale=\tikzscale]
      \begin{groupplot}[
        group style={
          group size=1 by 2,
          vertical sep=0cm,
          xticklabels at=edge bottom,
        },
        width=8cm,
        height=2.7cm,
        xmin=-0.8,
        xmax=6.8,
        xtick=data,
        xticklabels={Citus$_\mathit{SR}$,
        Citus$_\mathit{CJ}$,
        ShardingSphere$_\mathit{SR}$,
        ShardingSphere$_\mathit{CJ}$,
        Vitess$_\mathit{SR}$,
        Vitess$_\mathit{CJ}$,
        ClickHouse$_\mathit{SR}$
        },
        ybar=1pt,
        ylabel near ticks,
        /pgf/bar width=5pt,
        enlarge x limits=0.10,
        legend style={
          at={(0.4,1.25)},
          anchor=south,
          legend columns=2,
          draw=none,
          column sep=4pt
        }
      ]

      \nextgroupplot[
        font=\small,
        ymin=5,
        ymax=300,
        ytick={10,100,200,300},
        axis y discontinuity=parallel,
        axis x line=bottom,
        x axis line style={dotted,-},
        axis y line=left,
                        ylabel={Mult. factor ($\times$)},
        ylabel style={xshift=-0.55cm},
        %ylabel={\Large Mult. Factor ($\times$)},
        %ylabel style={xshift=-0.55cm},
        % ymode=log,
        % log basis y={10},
        % log ticks with fixed point,
        % scaled y ticks=false,
      ]
      \addplot[draw=blue,
        nodes near coords={\pgfmathprintnumber[fixed,precision=2]{\pgfplotspointmeta}},
        every node near coord/.append style={font=\small, yshift=0.5pt}
      ] coordinates {
        (0,1.0)
        (1,1.0)
        (2,1.0)
        (3,1.0)
        (4,1.0)
        (5,1.0)
        (6,1.0)
      };
      \addplot[draw=red, fill=red,
        nodes near coords={\pgfmathprintnumber[fixed,precision=2]{\pgfplotspointmeta}},
        every node near coord/.append style={font=\small, yshift=0.5pt}
      ] coordinates {
        (0,225.66)
        (1,3.80)
        (2,266.67)
        (3,198.83)
        (4,2.76)
        (5,97.52)
        (6,nan)
      };
      \legend{\large \mytool$_{\mathit{!MT}}$,\large \mytool}
\hspace{2ex}
      \nextgroupplot[
        font=\small,
        ymin=0,
        ymax=6,
        ytick={0,2,4},
        axis x line=bottom,
        axis y line=left,
        y axis line style={-},
        x tick label style={rotate=30, anchor=north east},
      ]
      \addplot[draw=blue,
        nodes near coords={\pgfmathprintnumber[fixed,precision=2]{\pgfplotspointmeta}},
        every node near coord/.append style={font=\small, yshift=0.5pt}
      ] coordinates {
        (0,1.0)
        (1,1.0)
        (2,1.0)
        (3,1.0)
        (4,1.0)
        (5,1.0)
        (6,1.0)
      };
      \addplot[
        draw=red,
        fill=red,
        point meta=y,
        nodes near coords={\pgfmathprintnumber[fixed,precision=2]{\pgfplotspointmeta}},
        every node near coord/.append style={font=\small, yshift=2pt}
      ] coordinates {
        (0,225.66)
        (1,3.80)
        (2,266.67)
        (3,198.83)
        (4,2.76)
        (5,97.52)
        (6,5.94)
      };

    \end{groupplot}
  \end{tikzpicture}
    \end{scaletikzpicturetowidth}
  \end{minipage}

   \captionsetup{skip=0pt}
  \caption{Frequency of triggering distributed query optimizations,  shard routing (SR) and co-located joins (CJ), across four DDBMSs.
  For ClickHouse, we evaluate only SR, as it does not support CJ.
 Results are normalized, and the y-axis reports multiplicative factors  relative to the baseline.  }
  \label{fig:trigger-rate-both}

\end{figure*}
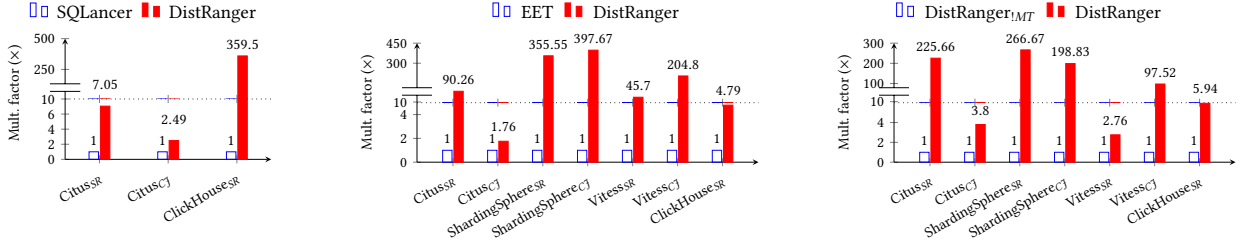

As shown in Figure~\ref{fig:trigger-rate-both} (left and middle), 
\mytool{} triggers the two distributed query optimizations  more frequently than both SQLancer and EET, 
achieving improvements of up to several hundred times. 
This is expected, as our mutation strategies, i.e.,
\(\Delta_{\mathit{PI}}\) and \(\Delta_{\mathit{JA}}\), 
are explicitly designed to trigger these optimizations, 
while query generation in SQLancer and EET 
is largely agnostic to them.

% Main-paper Citus pattern figure definitions.
% Requires: \usepackage{tikz}, \usepackage{pgfplots}, \usepgfplotslibrary{groupplots}
\providecommand{\patternlegendbox}[1]{\tikz[baseline=-0.6ex]{\draw[fill=#1,draw=#1] (0,0) rectangle (0.55em,0.55em);}}
\providecommand{\patternlegendboxopen}[1]{\tikz[baseline=-0.6ex]{\draw[draw=#1,line width=0.45pt,fill=white] (0,0) rectangle (0.55em,0.55em);}}
\providecommand{\drawaccessseparatorsfour}[1]{%
  \draw[densely dashed,gray!45,line width=0.45pt] (axis cs:1.5,0) -- (axis cs:1.5,#1);
  \draw[densely dashed,gray!45,line width=0.45pt] (axis cs:2.5,0) -- (axis cs:2.5,#1);
}
\providecommand{\drawaccessseparatorone}[1]{%
  \draw[densely dashed,gray!45,line width=0.45pt] (axis cs:1.5,0) -- (axis cs:1.5,#1);
}

\pgfplotsset{
  observedcituspattern/.style={
    ybar,
    bar width=2.4pt,
    width=0.615\columnwidth,
    height=0.315\columnwidth,
    ymin=0,
    xmin=0.80,
    xmax=4.20,
    enlarge x limits=false,
    xtick={1,2,3,4},
    xticklabels={Local,Reference,\underline{Shard routing},All-shard},
    tick label style={font=\scriptsize},
    y tick label style={font=\scriptsize},
    label style={font=\scriptsize},
    ylabel style={font=\scriptsize},
    title style={font=\scriptsize},
    x tick label style={rotate=25, anchor=east, font=\scriptsize},
    ymajorgrids=true,
    major grid style={dashed,gray!30},
    tick style={draw=none},
    ylabel near ticks,
    axis x line*=bottom,
    axis y line*=left,
  },
  observedcitusjoin/.style={
    ybar,
    bar width=2.2pt,
    width=0.435\columnwidth,
    height=0.315\columnwidth,
    ymin=0,
    symbolic x coords={Coloc.,Ref.,Distributed},
    xmin={[normalized]-0.20},
    xmax={[normalized]2.20},
    enlarge x limits=false,
    xtick=data,
    xticklabels={\underline{Co-located},Reference,Cross-shard},
    tick label style={font=\scriptsize},
    y tick label style={font=\scriptsize},
    label style={font=\scriptsize},
    ylabel style={font=\scriptsize},
    title style={font=\scriptsize},
    x tick label style={rotate=25, anchor=east, font=\scriptsize},
    ymajorgrids=true,
    major grid style={dashed,gray!30},
    tick style={draw=none},
    ylabel near ticks,
    axis x line*=bottom,
    axis y line*=left,
  },
  sqlancerbarsmall/.style={draw=blue, fill=white, line width=0.45pt, bar shift=-4.2pt},
  eetbarsmall/.style={draw=green!55!black, fill=white, line width=0.45pt, bar shift=-1.4pt},
  nomtbarsmall/.style={draw=brown, fill=white, line width=0.45pt, bar shift=1.4pt},
  distrangerbarsmall/.style={draw=red, fill=red, line width=0.45pt, bar shift=4.2pt},
}

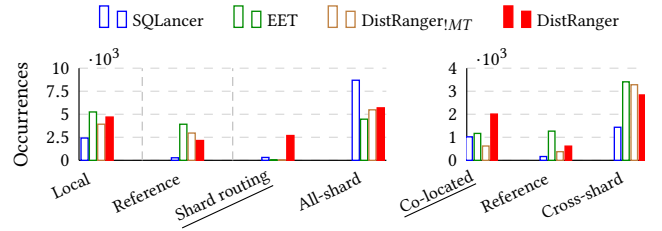
\begin{figure}
  \centering

  \resizebox{\columnwidth}{!}{%
  \begin{tikzpicture}
    \begin{groupplot}[
      group style={group size=2 by 1, horizontal sep=0.95cm},
    ]
      % -------- Citus shard-access --------
      \nextgroupplot[
        observedcituspattern,
        ymax=10000,
        ytick={0,2500,5000,7500,10000},
        scaled y ticks=base 10:-3,
        ylabel={\footnotesize Occurrences},
        legend to name=cituspatternlegend,
        legend columns=4,
        legend style={
          draw=none,
          fill=none,
          font=\scriptsize,
          cells={anchor=west},
          /tikz/every even column/.append style={column sep=0.28cm},
        },
      ]
      \addplot+[sqlancerbarsmall] coordinates {(1,2420) (2,278) (3,313) (4,8695)};
      \addlegendentry{SQLancer}

      \addplot+[eetbarsmall] coordinates {(1,5251) (2,3918) (3,28) (4,4467)};
      \addlegendentry{EET}

      \addplot+[nomtbarsmall] coordinates {(1,3921) (2,2961) (3,12) (4,5483)};
      \addlegendentry{\mytool{}$_{\mathit{!MT}}$}

      \addplot+[distrangerbarsmall] coordinates {(1,4698) (2,2153) (3,2694) (4,5697)};
      \addlegendentry{\mytool{}}

      \drawaccessseparatorsfour{10000}

      % -------- Citus distributed-join --------
      \nextgroupplot[
        observedcitusjoin,
        ymax=4000,
        ytick={0,1000,2000,3000,4000},
        scaled y ticks=base 10:-3,
        ylabel={},
      ]
      \addplot+[sqlancerbarsmall] coordinates {(Coloc.,1020) (Ref.,165) (Distributed,1436)};
      \addplot+[eetbarsmall] coordinates {(Coloc.,1169) (Ref.,1268) (Distributed,3409)};
      \addplot+[nomtbarsmall] coordinates {(Coloc.,622) (Ref.,377) (Distributed,3283)};
      \addplot+[distrangerbarsmall] coordinates {(Coloc.,2008) (Ref.,616) (Distributed,2839)};
    \end{groupplot}

    % Native PGFPlots bar legend, centered above both subplots.
    \node[anchor=south] at
      ($(group c1r1.north west)!0.5!(group c2r1.north east)+(0,2mm)$)
      {\pgfplotslegendfromname{cituspatternlegend}};
  \end{tikzpicture}%
  }

  \captionsetup{skip=0pt}
  \caption{Distributions of shard-access patterns (left), including local, reference-table, and distributed-table access, and distributed-join patterns (right), including co-located, reference-table, and cross-shard joins, exercised by 10K executable queries in Citus.
  The distributed-table access patterns include shard routing and all-shard access.}
  \label{fig:citus-patterns}
  \vspace{-2ex}
\end{figure}

\subsubsection{Coverage of Distributed Execution Patterns} 
To gain a more comprehensive understanding of how  queries generated by each tool exercise different distributed execution behaviors, we additionally measure various \emph{shard-access} and \emph{distributed-join} patterns
(Section~\ref{sec:Optimization}).
This provides complementary insights into DDBMS behaviors beyond the trigger counts reported earlier for
shard routing and co-located joins.

Due to space limitations, 
we present the results for Citus in Figure~\ref{fig:citus-patterns} 
and defer the results for the remaining DDBMSs to the Appendix~\cite{tech-rpt}.\footnote{For Citus, we analyze \emph{query plans}, which expose accessed tables, shards, and whether joins are executed in a co-located manner or across shards.}
All results align with our expectations:
for the two target distributed optimizations, namely shard routing and co-located joins, significantly more queries generated by \mytool{} exercise the corresponding patterns compared with SQLancer and EET.
Additionally,
due to the randomness in query generation, 
\mytool{} also exercises patterns beyond the two target optimizations, and each query may exercise zero or multiple patterns.

\subsection{A Closer Look At \mytool  }\label{sec:Sensitivity}
%Having demonstrated \mytool's effectiveness over state-of-the-art DBMS testing tools, 
%we now analyze how each of its key design choices contributes to these gains.

We now analyze how each of \mytool's key design choices contributes to its effectiveness 
over  state-of-the-art DBMS testers.

\subsubsection{Ablation Analysis} \label{sss:ablation}
We examine two variants: (i) \mytool{}  without query mutation 
 (i.e., using random query generation instead),
denoted as \mytool$_\mathit{!MT}$,
to assess its effect on triggering diverse distributed query execution and optimizations;
and 
(ii) \mytool{} without the feedback mechanism,
denoted as  \mytool$_\mathit{!FB}$,
to evaluate its impact on the executability of generated queries.

Overall, both variants perform significantly worse on their respective metrics,
as expected.
In particular, as shown in Table~\ref{tab:validity-compare}, \mytool$_\mathit{!FB}$ 
exhibits a  lower query executability rate 
than \mytool{} across the four DDBMSs,
e.g., with up to a 2.2$\times$ reduction for ShardingSphere. 
Moreover, \mytool$_\mathit{!MT}$ explores fewer distinct feature paths than \mytool{}
within the same time budget,
as shown in Figure~\ref{fig:diversity}, 
with up to a 6$\times$ reduction for ShardingSphere.

%Moreover, \mytool$_\mathit{!MT}$ 
%covers fewer unique feature paths than \mytool{} across all four DDBMSs, as shown in Figure~\ref{fig:diversity}.
%\mytool{} continuously explores new feature paths, while 
%\mytool$_\mathit{!MT}$ grows more slowly, e.g., with up to a 6$\times$ reduction for ShardingSphere.

%triggers distributed query optimizations  less frequently than \mytool.
%\nobi{For planners with stricter optimization preconditions}
%(e.g., ShardingSphere),
%this gap reaches on the order of  $10^2\times$;
%see Figure~\ref{fig:trigger-rate-both} (right).

%\todo{\mytool$_\mathit{!MT}$; reference Figure~\ref{fig:trigger-rate-both} }
%As shown in Figure~\ref{fig:trigger-rate-both}, \mytool$_\mathit{!MT}$ triggers both optimizations far less frequently than \mytool{}.
%For planners whose optimization preconditions are stricter (e.g., those in ShardingSphere), this gap reaches about 90--200$\times$.
%For more optimization-friendly planners, the gap is typically around 2.5--10$\times$.

%Another  advantage of our query mutation is its generality:
%it can be instantiated to target specific distributed query optimizations, such as shard routing and co-located joins (Section~\ref{subsec:mutation}).
%Accordingly,
In addition, we evaluate the effectiveness of our mutation strategies in targeting distributed query optimizations. 
Specifically, we compare how frequently \mytool{} and \mytool$_{\mathit{!MT}}$ trigger the two  optimizations, namely shard routing and co-located joins. 
%The experimental setup is the same as that used in Section~\ref{optimization_expriments}.
%For each tool, we generate 10K executable queries per run across five runs under randomly generated DDBMS instances.
%We instrument key entry points in the optimization logic to record when each optimization is triggered (see \cite[Appendix~\ref{x}]{tech-rpt} for details); 
%thus, trigger counts approximate its execution frequency. 
%For ClickHouse, we evaluate only shard routing, as it does not support co-located joins.
As shown in Figure~\ref{fig:trigger-rate-both} (right)
and Figure~\ref{fig:citus-patterns} 
(with additional results in the Appendix~\cite{tech-rpt}),
\mytool$_{\mathit{!MT}}$ triggers both optimizations much less frequently than \mytool.
%\footnote{The same observation holds  when comparing \mytool{} with SQLancer and EET; see \cite[Appendix~\ref{x}]{tech-rpt} for  the experimental results.} 
For instance, the gap 
in Figure~\ref{fig:trigger-rate-both}
reaches approximately 190--250$\times$ on ShardingSphere.
%For systems with stricter optimization preconditions (e.g., ShardingSphere),

Finally,
we  examine how the above degradations impact bug detection capability.
As shown in Table~\ref{tab:24hourBug-compare}, both variants find noticeably fewer bugs than \mytool.
This is expected: 
without the feedback mechanism, 
fewer queries can be successfully executed
by the DDBMSs in the first place; 
 without the query mutation, the probability of triggering diverse distributed execution behaviors is reduced, thereby lowering the likelihood of exposing  bugs.

\begin{figure}[t]
\centering
\begin{tikzpicture}
% left: Validity
\begin{axis}[
  name=main,
    tick style={draw=none},
  width=0.7\linewidth,
  height=3cm,
  xmin=0, xmax=30,
  ymin=50, ymax=90,
  ylabel={QE rate (\%)},
  axis y line*=left,
  %axis x line*=bottom,
  xtick={0,5,10,15,20,30},
  xticklabels={0,5,10,15,20,30},
  xlabel={$\theta$ (\%)},
          ytick={50,60,70,80,90},
    xticklabel style={font=\footnotesize},
     yticklabel style={font=\footnotesize},
       label style={font=\footnotesize}, 
  legend style={at={(1.4,.5)},anchor=south,legend columns=2,draw=none},
]
  \addplot[mark=*, thick, color=blue] coordinates {
    (0,58.8)
    (5,68.0)
    (10,79.6)
    (15,83.4)
    (20,82.0)
    (30,83.8)
  };
  \addlegendentry{ \small QE}
\end{axis}

% right: False-Positive Rate
\begin{axis}[
  width=0.7\linewidth,
      tick style={draw=none},
  height=3cm,
  at={(main.south west)},
  anchor=south west,
  xmin=0, xmax=30,
  ymin=0, ymax=30,
  ylabel={FP rate (\%)},
  axis y line*=right,
  axis x line=none,
  xtick=\empty,
  ymajorgrids=false,
      xticklabel style={font=\footnotesize},
     yticklabel style={font=\footnotesize},
            label style={font=\footnotesize}, 
  legend style={at={(1.4,.2)},anchor=south,legend columns=2,draw=none},
]
  \addplot+[mark=square, thick, color=red] coordinates {
    (0,2.4)
    (5,2.9)
    (10,5.8)
    (15,7.9)
    (20,10.5)
    (30,22.7)
  };
  \addlegendentry{ \small FP}
\end{axis}
\end{tikzpicture}
   \captionsetup{skip=0pt}
\caption{Query executability (QE) and false-positive (FP) rates under different values of $\theta$.}
\label{fig:theta}
\vspace{-2ex}
\end{figure}
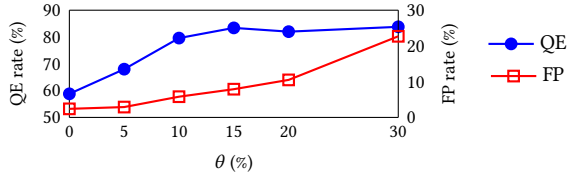

% Required packages:
% \usepackage{booktabs}
% \usepackage{tikz}
% \usepackage{pgf-pie}
% \usepackage{xcolor}
% \usepackage{array}
% \usepackage{subcaption}
% \captionsetup{subrefformat=parens}
%
% Replace the four TBD values below with the measured component sizes.
\newcommand{\memSDM}{0.84 MB}
\newcommand{\memAST}{0.03 MB}
\newcommand{\memPathSet}{615.98 MB}
\newcommand{\memOthers}{24.30 MB}
\newcommand{\memTotal}{641.15 MB}

\definecolor{stateblue}{RGB}{70,120,170}
\definecolor{execorange}{RGB}{255,132,20}
\definecolor{overheadpurple}{RGB}{177,119,165}
\definecolor{detailgray}{RGB}{220,220,220}
\definecolor{astgreen}{RGB}{82,160,75}
\definecolor{mutred}{RGB}{230,80,80}
\definecolor{searchcyan}{RGB}{120,190,190}
\definecolor{updateyellow}{RGB}{245,210,50}

\begin{figure*}[t]
\centering

% ---------------- (a) Overall memory ----------------
\begin{minipage}[c]{0.2\textwidth}
\centering
\phantomsubcaption\label{fig:memory-overhead-overall}
{\small
\renewcommand{\arraystretch}{1.0}
\setlength{\tabcolsep}{2.5pt}
\begin{tabular*}{\linewidth}{@{\extracolsep{\fill}}lr@{}}
\toprule
\textbf{Tool} & \textbf{Memory} \\
\midrule
SQLancer   & 269.06 MB \\
EET        & 12.19 MB \\
DistRanger & 641.15 MB \\
\bottomrule
\end{tabular*}
}

\vspace{6pt}
{\small (a) total memory usage comparison}
\end{minipage}
\hspace{4ex}
% ---------------- (b) Memory breakdown ----------------
\begin{minipage}[c]{0.2\textwidth}
\centering
\phantomsubcaption\label{fig:memory-overhead-components}
{\small
\renewcommand{\arraystretch}{1.0}
\setlength{\tabcolsep}{2.5pt}
\begin{tabular*}{\linewidth}{@{\extracolsep{\fill}}lr@{}}
\toprule
\textbf{Component} & \textbf{Memory} \\
\midrule
SDM              & \memSDM     \\
Query ASTs       & \memAST     \\
Feature paths    & \memPathSet \\
Others           & \memOthers  \\
%\midrule
%\textbf{Total}  & \textbf{\memTotal} \\
\bottomrule
\end{tabular*}
}

\vspace{6pt}
{\small (b) memory breakdown of \mytool}
\end{minipage}
\hfill
% ---------------- (c) Runtime breakdown ----------------
\begin{minipage}[c]{0.55\textwidth}
\centering
\phantomsubcaption\label{fig:memory-overhead-runtime}
\footnotesize

% Legend. The vertical rule separates the main-pie components
% from the detailed components of the 0.91% slice.
\begin{tabular}{@{}ll!{\color{gray!60}\vrule width 0.6pt}ll@{}}
\textcolor{stateblue}{\rule{0.65em}{0.48em}}~DDBMS gen. &
\textcolor{overheadpurple}{\rule{0.65em}{0.48em}}~Other runtime overhead &
\textcolor{astgreen}{\rule{0.65em}{0.48em}}~Query gen. &
\textcolor{searchcyan}{\rule{0.65em}{0.48em}}~Maintain path set \\
\textcolor{execorange}{\rule{0.65em}{0.48em}}~Query exec. &
\textcolor{detailgray}{\rule{0.65em}{0.48em}}~RHS components &
\textcolor{mutred}{\rule{0.65em}{0.48em}}~Mutation &
% \textcolor{updateyellow}{\rule{0.65em}{0.48em}}~Updating path set
\end{tabular}

%\vspace{1pt}
\resizebox{0.8\columnwidth}{!}{%
\begin{tikzpicture}

% Main pie
\node at (-2.3,0) {
\begin{tikzpicture}
\pie[
    radius=0.7,
    text=inside,
    hide number,
    color={stateblue,execorange,overheadpurple,detailgray},
    sum=100
]{
    12.59/,
    54.37/,
    32.13/,
    1.91/
}

% Manually place only the labels you want to keep
\node[font=\scriptsize] at (16:0.43) {6.59\%};
\node[font=\scriptsize] at (150:0.3) {58.37\%};
\node[font=\scriptsize] at (300:0.35) {34.13\%};

% No label for 0.91%
\end{tikzpicture}
};

% ---------------- Arrow ----------------
% Slightly longer
\draw[->, >=stealth, line width=0.9pt, gray]
    (-1.30,-0.02) -- (0.85,-0.02);

% Optional label for zoomed part
\node[font=\scriptsize] at (-0.9, 0.12) {0.91\%};

% Detail pie
\node at (1.9,0) {
\begin{tikzpicture}
\pie[
    radius=0.7,
    text=inside,
    color={astgreen,mutred,searchcyan},
    hide number,
    sum=0.91,
    every label/.style={font=\scriptsize},
    every number/.style={font=\scriptsize}
]{
    0.11/,
    0.43/,
    0.37/
}
\node[font=\scriptsize] at (14:0.42) {0.11\%};
\node[font=\scriptsize] at (150:0.35) {0.43\%};
\node[font=\scriptsize] at (300:0.4) {0.37\%};

\end{tikzpicture}
};

\end{tikzpicture}%
}

%\vspace{2pt}
{\small (c) runtime breakdown of \mytool}
\end{minipage}
  \captionsetup{skip=5pt}
\caption{Memory usage and runtime breakdown during 24-hour testing of Vitess.
Other memory usage and runtime overhead are mainly attributed to the logging and checkpointing mechanisms
used for bug probing.}
\label{fig:memory-runtime-overhead}

\end{figure*}
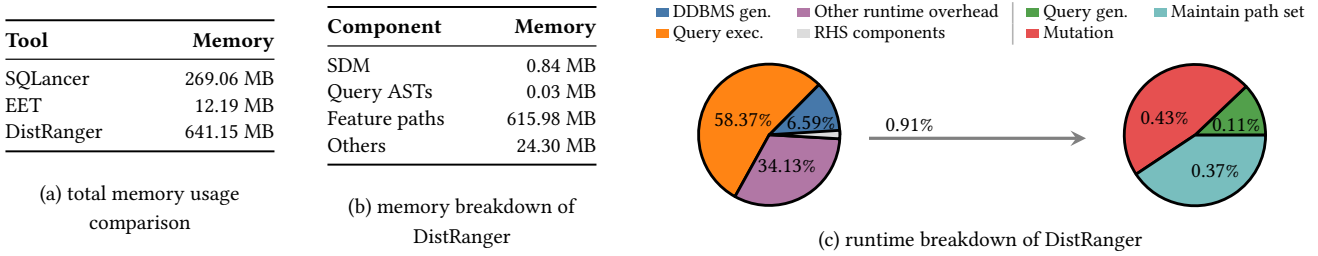

\subsubsection{Threshold Selection and Trade-offs} \label{sss:theta}
In our approach, the feedback mechanism rejects a feature path, 
and thus any query containing it,
when its 
historical query executability (QE) rate falls below $\theta$ (Section~\ref{sec:feedback}).
Figure~\ref{fig:theta} shows our analysis of 40K queries (10K per DDBMS). 
As $\theta$ increases, the overall QE rate improves and eventually stabilizes, while the false-positive (FP) rate continues to increase.
The FP rate is defined as the fraction of executable queries that are  rejected by the feedback filter.
%albeit at a decreasing rate.

In black-box testing, false positives are difficult 
%if not impossible,
to avoid.
% to avoid, as $\phi$ cannot fully capture the behaviors of DDBMS optimizers.
%\zz{to avoid, as normalization and annotation cannot fully capture all the behaviors of DDBMS optimizers.}
Consider a simple example: a predicate such as \sqlkw{WHERE} \sqlkw{TRUE} \sqlkw{OR} \textit{right} always evaluates to true.
The optimizer may prune the \textit{right} branch 
as a simplification, 
even if it contains  unsupported %structural 
features. 
As a result, the query can be executed by the DDBMS. 
When $\theta$ is too small
(e.g., 0\%--5\%),
 branch pruning can delay learning, as 
the feedback mechanism
 may require many counterexamples before recognizing that a query is actually not executable, resulting in a limited QE rate. 
When $\theta$ is too large (e.g.,  $\geq 20\%$),
executable feature paths may be misclassified as unsupported by our feedback mechanism,
leading to higher FP
rates and overly conservative mutations. 
Hence, in practice, we use a moderate $\theta$ (e.g., 10\%) to balance this trade-off.

\subsection{Resource Overhead Analysis} \label{sec:overhead}
Analyzing the resource costs of a testing tool provides valuable insights into its practical trade-offs.
Therefore, we measure the memory usage and runtime overhead of \mytool{} 
using a 24-hour testing campaign on Vitess as a case study. 
%Specifically, we analyze the contributions of different components to the overall resource consumption and compare \mytool{} with its competitors.

First,  the primary source of \mytool{}'s memory usage is storing the feature path set,
as shown in Figure~\ref{fig:memory-overhead-components}.
In comparison, the overhead of storing the SDM and query ASTs is negligible. This also explains why \mytool{} consumes more memory than SQLancer and EET, as shown in Figure~\ref{fig:memory-overhead-overall}. 
Nevertheless, this additional memory overhead is a worthwhile trade-off, since the feature path set underlies our feedback mechanism, which improves query executability and thus enhances bug-finding effectiveness, 
as we have demonstrated.
Note that although \mytool{} incurs higher memory usage, its overall memory consumption during 24-hour testing remains practical on modern machines.

Moreover, Figure~\ref{fig:memory-overhead-runtime}  shows the runtime breakdown of \mytool{}.
%, including the cost of maintaining the feature path set. 
The majority of  runtime is spent on 
executing queries
 and other components, 
 including
 the
 logging and checkpointing mechanisms used for bug probing. 
In contrast, maintaining the feature path set accounts for only 0.37\% of the total runtime overhead, indicating that this cost is negligible.

\section{Related Work } 
\label{sec:related}
\inlsec{(D)DBMS Testing}
Recent years have witnessed a surge of efforts on testing DBMSs,
including 
SQLancer~\cite{SQLancer,PQS} and EET~\cite{EET}.
However, 
bugs arising from query processing and optimization in distributed settings remain largely underexplored.
Addressing this gap is the focus of our approach \ourapproach.

%SQLancer~\cite{SQLancer,PQS} and EET~\cite{EET}
%are state-of-the-art tools  integrating both test generation
%(e.g., queries and database schemas)
%and test oracles. 
%Despite their success in finding bugs in production DBMSs, they focus on centralized settings.
%In contrast,  \ourapproach  is specifically designed for testing DDBMSs, targeting query-related issues in the distribution layer  that are absent in centralized DBMSs.
%In particular, our experiments demonstrate that \ourapproach outperforms both approaches in bug detection, with nearly all discovered bugs being specific to distributed query processing.

DistSQL~\cite{10.1145/3786673} is a recent differential testing framework that targets logic bugs in DDBMSs.
It compares the results of the same  query executed on a DBMS configured in centralized and distributed modes.
Yet, its code is not publicly available as of this writing.
%\footnote{\textcolor{NavyBlue}{We also contacted the authors to request the code but received no response.}} 
Given its complexity (approximately 28 KLOC),
%and insufficient implementation details in the paper,
a  reimplementation 
 would require substantial assumptions and undermine a fair experimental comparison.
 %\footnote{\textcolor{NavyBlue}{For example, the paper does not specify the distribution strategies used for initializing the DDBMSs or how queries are generated with respect to these strategies.  Moreover, it provides insufficient details for implementing concrete DDBMS state mutators, e.g., how to add, delete, or alter tables.}}
Hence,
we provide below 
a qualitative comparison.
%based on the information available in the paper.

Overall, \ourapproach
differs from DistSQL in three key aspects:

\begin{enumerate}[leftmargin=18pt]
    \item While \ourapproach currently uses the same oracle for  identifying bugs,
    it is designed as a  general testing framework.
    Additional test oracles can be 
naturally incorporated to enable the detection of a broader range of issues (see below and Section~\ref{sec:concl}).
In contrast, DistSQL tightly couples its logic bug detection with differential testing, 
relying on query plan discrepancies across deployment modes 
to generate queries that are more likely to trigger bugs.
%\zz{, and further uses query plan, including plan discrepancies across the two deployment modes, to explore better query seeds.}

    \item \ourapproach emphasizes generating executable test queries, whereas DistSQL does not. Therefore, we conjecture that a     non-trivial fraction  of DistSQL-generated queries may be rejected by DDBMSs, thereby reducing overall testing efficacy.
    
    \item \ourapproach enables targeted generation of queries that trigger specific distributed optimizations via SDM-guided mutation (i.e., $\mu$),
whereas DistSQL lacks such targeted exploration, potentially missing bugs related to these optimizations.
    %Moreover, the bugs it discovers tend to be general logic bugs and may be less informative.

\end{enumerate}

Nonetheless, DistSQL can  expose certain classes of bugs  not yet covered by \mytool. 
First, its query generator supports SQL features not emphasized in our implementation (e.g., window functions with \sqlkw{PARTITION BY}), enabling it to generate relevant bug-triggering queries.
Second, it explores bugs related to system configurations, such as %varying storage engines or 
enabling optional components
or changing parameters.

These gaps reflect implementation choices rather than fundamental limitations of \ourapproach.
%Missing SQL features can be incorporated into the base query generator without altering the core design, after which the mutation strategy can be extended and the feedback mechanism can continue to guide it using executability results. 
%Configuration-level testing requires additional engineering effort, as noted in DistSQL. 
%These extensions are left for future work.
To demonstrate its extensibility,
we performed a preliminary extension of \mytool{} 
by incorporating 
ShardingSphere's  
configuration options
 into the DDBMS instance generation stage.
With this extension, we uncovered a bug 
in
ShardingSphere,
which has been confirmed by the developers.
This bug is triggered when two configurations, i.e., encryption and auto key-generation, are enabled.
Compared with the setting where both options are disabled,
the same query produces different results.

There is also a substantial body of research on testing distributed systems. 
While some   involves distributed databases, its focus is largely 
orthogonal,
e.g., on synchronization failures~\cite{depstate} and
recovery bugs~\cite{CrashFuzz}.
%, and 
%data inconsistency~\cite{jepsen}.
 % security issues~\cite{mallory}.
In contrast, \ourapproach
%, as well as SQLancer and DistSQL, 
focuses on issues arising from  query processing.
Exploring synergies between \ourapproach and these techniques, e.g., fault injection~\cite{jepsen,fis}, is  interesting  future work,
as it may, e.g., 
help trigger more diverse distributed execution behaviors.

\smallskip
From a technical perspective, our work is closely related to two lines of DBMS testing research, i.e.,
test generation and test oracles,
which have thus far primarily focused on centralized settings.

\inlsec{Test Generation}
A large body of prior work~\cite{SQLsmith,dynsql,Mozi,go-randgen,lego,QPG,sqlright,TQS} 
focuses on generating diverse test cases,
including complex SQL queries and varied database schemas and configurations. 
\ourapproach builds upon these advances, e.g., DynSQL~\cite{dynsql}.
Among these approaches, mutation-based techniques are often used to increase the diversity of queries, %~\cite{griffin,sqlright}, 
while feedback mechanisms %~\cite{QPG,lego} 
are employed to guide test generation, aiming to improve test case quality (e.g., trigger a wider range of system behaviors).

However, queries that are syntactically valid and accepted by centralized DBMSs are not necessarily executable in DDBMSs due to distribution-specific constraints.
Paradoxically, 
increasing query complexity, which has proven effective in triggering diverse system behaviors in centralized settings, can  raise the likelihood of rejection in distributed settings. 
\ourapproach addresses this dilemma 
by  
mutating queries based on the SDM 
and
leveraging historical executability results to  guide the mutation process.

\inlsec{Test Oracles}
An orthogonal line of work
is dedicated to the design of test oracles,
 which determine whether a test output matches the expected result.
Representative oracles include
(i) differential testing~\cite{SQLsmith,go-randgen}, which compares outputs of identical queries executed under different configurations;
%(e.g., centralized vs. distributed deployments of the same DBMS in our experiments);
(ii) metamorphic testing~\cite{TLP,DQP,EET,constant,srs,10.14778/3734839.3734861,10.14778/3659437.3659445},
which verifies expected relations among the outputs of related inputs;
%(e.g., SQLancer's and EET's test oracles described in Section~\ref{sec:Comparison});
and 
(iii) crash and exception detection~\cite{dynsql}, achievable by simply inspecting system outputs.

\ourapproach is designed as a general testing framework that currently focuses on test generation and already integrates the oracles in (i) and (iii) within \mytool. 
Extending it to incorporate additional oracles, particularly those in (ii), is feasible, 
either directly or with minor adaptations.
For instance, incorporating SQLancer's TLP
or        EET's semantically equivalent query transformations
incurs no additional cost:
for a  query  and its derived query,
we simply obtain an additional pair of execution results in the distributed setting;
as a result, we compare the equivalence of four query results in total. 
\section{Discussion and Conclusion
}
\label{sec:concl}
We have presented \ourapproach, a novel automated testing approach tailored to DDBMSs, 
together with its implementation, \mytool.
Using \mytool, we have uncovered many previously unknown bugs in production DDBMSs
%(31 in total, with 21 confirmed and 15 fixed to date)
and demonstrated its superior effectiveness in bug detection compared to the state-of-the-art. 
%\ourapproach can be instantiated and extended to test a wide range of DDBMSs, as well as distributed query processing and optimization techniques (see below).
This work complements prior efforts  on centralized settings by addressing issues  that arise specifically in the distribution layers of DDBMSs.
% Together, these efforts contribute to more reliable database systems.

%\nobi{ \ourapproach models table-level Data-Distribution Strategies (DDSs) to generate richer DDBMS states, so that generated SQL queries can trigger more DDS-related optimizations and fewer DDS-related limitations. 
%Using \mytool, we discovered 31 previously unknown distribution-related bugs on four DDBMSs, and outperformed state-of-the-art testing tools in our evaluation.}

Next,
we discuss the limitations of this work and outline additional directions for future work beyond those covered in Section~\ref{sec:related}.

\inlsec{Limitations}
While testing can find bugs, it does not guarantee their absence. 
\ourapproach
%, like other bug detection approaches,
shares this inherent limitation. 
Formally verifying  DDBMSs could overcome this issue,
but it is challenging~\cite{10.1145/3477132.3483540,DBLP:conf/nsdi/HowardKACC25}. %, if not infeasible, for large-scale distributed systems~ 
Moreover, \mytool{} may produce duplicate bug reports, as different test cases can trigger bugs with the same root cause.
We currently perform manual deduplication,
while
its automation 
remains a challenge in
system testing~\cite{DBLP:conf/pldi/ChenGZWFER13,10.1145/2771783.2771797}.
%We currently rely on manual deduplication; 
%yet its automation (and also for root cause diagnosis)
%is an orthogonal %and challenging
%research problem in system testing~\cite{DBLP:conf/pldi/ChenGZWFER13,10.1145/2771783.2771797}. 
Finally, \ourapproach
 relies on 
 explicit specifications to define how data is distributed.
 %, which are supported by many DDBMSs.
For systems with distribution strategies that are dynamically determined or not directly controllable, we could
extract
the realized data distribution from system-level metadata (e.g., region-level placement information in TiDB~\cite{tidb_region_placement})
and incorporate it into the testing process.

\inlsec{Extension to Other Distributed Query  Optimizations}
We have demonstrated \ourapproach's effectiveness in triggering distributed query optimizations
via shard routing and co-located joins. 
%Yet, DDBMSs support many 
Other optimizations
can be incorporated into \mytool{} by defining $\Delta$ rules, along with the mapping $\rho$, that transform the query to satisfy optimization-specific structural preconditions. 
For example,
testing 
\emph{aggregation pushdown}~\cite{citus_aggregate_functions,ss_group_by_merger} (e.g., \sqlkw{GROUP BY} on the sharding key) can be enabled by replacing tables with sharded counterparts,
rewriting the query to enforce the required aggregate and \sqlkw{GROUP BY} structure, and aligning grouping keys with sharding keys.

\inlsec{Discovering Other Types of Bugs}
\ourapproach can serve as a basis for detecting   other types of bugs,
e.g., 
%For example, it can be utilized to detect performance issues in DDBMSs, such as 
inefficient distributed query processing.
A key challenge  lies in the lack of effective test oracles, i.e., determining the expected execution time of a distributed query.
%Cardinality estimation~\cite{NoREC} may provide a promising direction for addressing it.  
Moreover, despite recent  advances in black-box checking of isolation levels~\cite{cobra,polysi,plume,elle,veristrong,DBLP:conf/icde/WeiXYLYCP25,10.1145/3742465}, e.g.,
 serializability and weaker levels~\cite{noc-noc,si},
they are largely limited to  key-value stores.
\ourapproach
can be extended 
to 
\emph{relational} DDBMSs by wrapping mutated SQL queries as transactions 
and exploiting their interactions with diverse distribution strategies to explore concurrency behaviors.
%leveraging existing oracles~\cite{adya1999weak,DBLP:conf/cav/BouajjaniER25} to validate their adherence to the claimed isolation guarantees.

%Notably,
%all these applications build on a common foundation:
%query executability, a prerequisite for testing to proceed in DDBMSs, 
%and 
%test diversity, an enabler of diverse system behaviors,
%both of which  lie at the core  of our \ourapproach approach.

\section*{ACKNOWLEDGMENTS} 
We appreciate the anonymous reviewers for their valuable feedback.
This work was partially supported by Shanghai Trusted Industry Internet Software Collaborative Innovation Center.
Si Liu was supported by the University Development Fund (UDF01004607).
Hengfeng Wei was supported by the
NSFC (62472214)
and
the Fundamental Research Funds for the Central Universities.

%\clearpage
\balance
\bibliographystyle{ACM-Reference-Format}
\bibliography{ref}

\appendix

\begin{figure}
\centering
\begin{minipage}{.95\linewidth}

\begin{lstlisting}[style=sql, firstnumber=1]
CREATE TABLE t1 (vkey INT4);
CREATE TABLE t2 (vkey INT4);
INSERT INTO t2 (vkey) values (5);
SELECT create_distributed_table('t1', 'vkey');    --- t1's DS
SELECT create_reference_table('t2');              --- t2's DS

SELECT * FROM (t2 FULL OUTER JOIN t1 ON (t2.vkey = t1.vkey))
WHERE NOT((85) IN (SELECT 1 FROM t2));
--- expected result: 1 row returned
--- actual result: 32 rows returned (*@\faBug@*) 
\end{lstlisting}

\end{minipage}
  \captionsetup{skip=5pt}
\caption{A logic bug found in Citus: \sqlkw{FULL OUTER JOIN} with a reference table yields 32 duplicate rows.}
\label{fig:bug2}
\end{figure}

% \begin{figure}[t]
%   \centering
%   \includegraphics[width=0.8\columnwidth]{figures/bug2.pdf}
%   \caption{A logic bug found in Citus: \sqlkw{FULL OUTER JOIN} with a reference table yields 32 duplicate rows.}
%   \label{fig:bug2}
% \end{figure}

\begin{figure}
\centering
\begin{minipage}{0.95\linewidth}

\begin{lstlisting}[style=sql, firstnumber=1]
SET DEFAULT SINGLE TABLE STORAGE UNIT = ds_0;     --- t0's DS
CREATE TABLE t0 (c2 TEXT,c6 TEXT);
SET DEFAULT SINGLE TABLE STORAGE UNIT = ds_1;     --- t1's DS
CREATE TABLE t1 (vkey INT4);
CREATE BROADCAST TABLE RULE t2;                   --- t2's DS
CREATE TABLE t2 (vkey INT4);

SELECT * FROM t0 AS ref_0
WHERE (ref_0.c2) IN (SELECT ref_0.c6 AS c_0
  FROM (t2 AS ref_1 FULL OUTER JOIN t1 AS ref_3
    ON (ref_1.vkey = ref_3.vkey)));
--- expected: the execution succeeds with correct result
--- actual result: the request times out after 30s (*@\faBug@*) 
\end{lstlisting}
%ERROR: HikariPool-27 - Connection is not available, request timed out after 30000ms 

\end{minipage}
  \captionsetup{skip=5pt}
\caption{A bug found in ShardingSphere: a correlated subquery triggers per-row rescans, exhausting the connection pool and causing timeouts.}
\label{fig:bug3}
\end{figure}

% \begin{figure}[t]
%   \centering
%   \includegraphics[width=0.8\columnwidth]{figures/bug3.pdf}
%   \caption{A bug found in ShardingSphere: Correlated subquery causes per-row re-scans and exhausts the connection pool, leading to timeout.}
%   \label{fig:bug3}
% \end{figure}

% \todo{to show another bug?}

% \begin{figure}
%     \centering
%     \input{figures/cumulative_bug_plot/cumulative_bug_plot}
%      \captionsetup{skip=2pt}
%     \caption{\reviewrightcap{R2}{4}{Maroon}{Cumulative  bugs  detected in the four DDBMSs
% over 24 hrs.
% Distribution-specific bugs are shown in parentheses.}}
%     \label{fig:cumulative_bugs2}
    
% \end{figure}

\section{More Experimental Results}
This section provides additional details to complement the experimental results in Section~\ref{sec:evaluation}.
It consists of five parts: 

\begin{enumerate}[leftmargin=15pt]

    \item a complete bug list, 
including detailed descriptions, bug types,  status, 
and SQL features involved in bug-triggering queries, as shown in Table~\ref{tab:bugs},
which corresponds to
 Table~\ref{tab:bug};
 %, Section~\ref{sec:DiscoverBugs}; 

\item 
two additional representative bugs 
found in Citus (Figure~\ref{fig:bug2})
and ShardingSphere
(Figure~\ref{fig:bug3}), respectively,
which
complement the one shown in Section~\ref{intro};

\item the
distribution of detected bugs by type,
as shown in Table~\ref{tab:bug-types}; 

\item the  entry points instrumented in the codebase of each DDBMS 
for measuring optimization triggering,
as shown in Table~\ref{tab:opt-trigger-entry-points};
%which
%corresponds to Section~\ref{optimization_expriments}.

\item the coverages of shard-access  (Figure~\ref{fig:shard-access-patterns2})
and distributed-join   (Figure~\ref{fig:distributed-join-patterns2})  patterns
for  ShardingSphere, Vitess, and ClickHouse.
\end{enumerate}

Below, we provide further details on (2), (4), and (5).

\subsection{Representative Bugs}

\begin{example}
Figure~\ref{fig:bug2} shows a SQL query that triggers a logic bug in Citus. The query performs a \sqlkw{FULL OUTER JOIN} between a distributed table \sqlvar{t1}, sharded by \sqlvar{vkey} (32 shards by default), and a reference table \sqlvar{t2}. Executing the query should return a single row. Instead, Citus returns 32 identical rows, one per shard, indicating that the full outer join is executed independently on each shard, with the coordinator simply concatenating the results rather than properly aggregating them.

This bug stems from a flaw in the distributed planner that prevents Citus from recognizing certain joins, e.g., \sqlkw{FULL OUTER JOIN} with pseudo-constant restrictions. 
As a result, the outer join is not considered during planning, leading to an incorrect distributed execution plan.
Citus treats this as a critical logic bug requiring user attention. In the current release, it explicitly raises an error instead of silently producing incorrect results.

%\bigskip
%Figure~\ref{fig:bug2} shows an incorrect query result from a distributed table \sqlkw{FULL OUTER JOIN} a reference table in Citus, where \sqlvar{t1} is distributed by \sqlvar{vkey} (32 shards by default) and \sqlvar{t2} is a reference table. Executing the query should return a single row.
% the unmatched row from \sqlvar{t2} joined with \sqlkw{NULL}s from \sqlvar{t1}, because the \sqlkw{IN} subquery evaluates to \sqlkw{TRUE}. 
%Instead, the system returns 32 identical rows—one per shard—indicating that the full outer join was effectively executed independently on all shards and the coordinator concatenated, rather than properly aggregating, the shard results.

%According to the maintainers, a distributed planner-hook visibility gap prevents Citus from seeing certain joins (e.g., \sqlkw{FULL OUTER JOIN} with pseudo-constant restrictions). As a consequence, Citus does not see the outer join during planning and produces an incorrect distributed plan. 

% \todo{zz: significance}
%This bug is particularly significant because it silently violates the semantics of a core SQL operator: instead of returning one row, the distributed execution returns one duplicate row per shard. 
%Moreover, Citus 13.2 officially treats this class of bugs as a significant wrong-result issue requiring user attention, and explicitly prefers throwing an error over silently producing incorrect results.
% The bug is fixed by PostgreSQL restoring unconditional invocation of the hook. 
\end{example}

 \begin{example}
Figure~\ref{fig:bug3} illustrates a query that exposes a timeout issue in ShardingSphere caused by connection pool exhaustion. 
The federation layer in ShardingSphere relies on SQL rewriting to retrieve data across data nodes,
 but it mishandles the correlated subquery. 
At runtime, for each outer row of \sqlvar{ref\_0}, it rescans all shards of \sqlvar{ref\_1} and \sqlvar{ref\_3} without reusing intermediate results. Moreover, each scan acquires a new backend connection, quickly exhausting the connection pool
and resulting in timeouts.

This bug is severe, as it can render the system unavailable:
once the connection pool is saturated, subsequent requests may be delayed or fail while waiting for available connections.
Moreover, it can be exploited by an attacker to launch  denial-of-service attacks. 
 \end{example}

\begin{table}
\captionsetup{skip=5pt}
  \caption{Distribution of the detected bugs by type.}
  \label{tab:bug-types}
 % \tablesize
  \centering
  \resizebox{\columnwidth}{!}{
  \begin{tabular}{l|cccc|c}
  %{p{1.2cm} C{0.7cm} C{1.5cm} C{0.7cm} C{1.2cm} |C{0.7cm}}
    \toprule
    \textbf{Bug Type} & \textbf{Citus} & \textbf{ShardingSphere} & \textbf{Vitess} & \textbf{ClickHouse} & \textbf{Total (31)} \\
    \midrule
    Logic            & 5 & 4 & 6 & 1 & 16 \\
    Crash            & 3 & 0 & 2 & 0 & 5 \\
    Error            & 0 & 3 & 3 & 0 & 6 \\
    Exception        & 0 & 1 & 0 & 1 & 2 \\
    Timeout          & 0 & 1 & 1 & 0 & 2 \\
    %\midrule
    %\textbf{Total}            & 7 & 10 & 12 & 2 & 31 \\
    \bottomrule
  \end{tabular}
  }
\end{table}

\begin{table}
\captionsetup{skip=5pt}
  \caption{Instrumented entry points for measuring optimization triggering.
  Each entry point corresponds to a specific function or source file serving as a measurement target.}
  \label{tab:opt-trigger-entry-points}
  \centering
  \resizebox{\columnwidth}{!}{
  \begin{tabular}{l l l}
    \toprule
    \textbf{DDBMS} & \textbf{Opt.} & \textbf{Instrumented entry point} \\
    \midrule
    Citus & SR & \texttt{shard\_pruning.c} \\
    Citus & CJ & \texttt{query\_colocation\_checker.c} \\
    ShardingSphere & SR & \texttt{routeByShardingConditions} \\
    ShardingSphere & CJ & \texttt{routeByShardingConditionsWithCondition} \\
    Vitess & SR & \texttt{RoutingParameters.equal} \\
    Vitess & CJ & \texttt{joinMerging.mergeShardedRouting} \\
    ClickHouse & SR & \texttt{optimize\_skip\_unused\_shards} \\
    \bottomrule
  \end{tabular}
  }
\end{table}

\subsection{Instrumented Entry Points}
To measure how frequently  generated queries trigger distributed query optimizations, we instrument entry points for the distributed query planning or routing logic in the codebase.
Table~\ref{tab:opt-trigger-entry-points} summarizes the instrumentation targets for the four DDBMSs.
Since different systems expose shard routing and co-located joins through different internal modules, the instrumentation is  system-specific. 

For Citus, we instrument its distributed planner. 
Specifically, we use \texttt{shard\_pruning.c} to measure shard-routing behavior because this file implements the shard routing (or pruning) logic that determines whether only a subset of shards needs to be accessed. 
For co-located joins, we instrument \texttt{query\_colocation\_checker.c}, which checks whether distributed tables satisfy the co-location requirements needed for co-located join execution.

For ShardingSphere, we instrument the Java routing layer and record whether execution reaches the corresponding route-engine logic inside the Java virtual machine. 
We use \texttt{routeBySharding}-\texttt{Conditions} to measure shard routing, since this component routes a query based on extracted sharding conditions. 
For co-located joins, we use \texttt{routeByShardingConditionsWithCondition}, which additionally handles routing decisions involving co-located joins.

For Vitess, we leverage its built-in query planning and routing components. 
We instrument \texttt{RoutingParameters.equal} to find when Vitess recognizes shard routing. 
For co-located joins, we instrument \texttt{joinMerging.mergeShardedRouting}, which is invoked when Vitess attempts to perform co-located joins.

For ClickHouse, we evaluate \texttt{optimize\_skip\_unused\_shards} by comparing query execution with this option enabled and disabled. 
A query is counted as triggering shard routing when enabling this option changes the internal execution path or query plan, indicating that ClickHouse can skip irrelevant shards by shard routing. 
ClickHouse does not expose the co-located join optimization.

\providecommand{\patternlegendbox}[1]{\tikz[baseline=-0.6ex]{\draw[fill=#1,draw=#1] (0,0) rectangle (0.55em,0.55em);}}
\providecommand{\patternlegendboxopen}[1]{\tikz[baseline=-0.6ex]{\draw[draw=#1,line width=0.45pt,fill=white] (0,0) rectangle (0.55em,0.55em);}}
\providecommand{\drawaccessseparatorsfour}[1]{%
  \draw[densely dashed,gray!45,line width=0.45pt] (axis cs:1.5,0) -- (axis cs:1.5,#1);
  \draw[densely dashed,gray!45,line width=0.45pt] (axis cs:2.5,0) -- (axis cs:2.5,#1);
}
\providecommand{\drawaccessseparatorone}[1]{%
  \draw[densely dashed,gray!45,line width=0.45pt] (axis cs:1.5,0) -- (axis cs:1.5,#1);
}

\pgfplotsset{
  observedpattern/.style={
    ybar,
    bar width=3.6pt,
    width=5.25cm,
    height=3.2cm,
    ymin=0,
    xmin=0.5,
    xmax=4.5,
    xtick={1,2,3,4},
    xticklabels={Local,Reference,\underline{Shard routing},All-shard},
    tick label style={font=\scriptsize},
    label style={font=\scriptsize},
    title style={font=\small},
    x tick label style={rotate=28, anchor=east, font=\scriptsize},
    ymajorgrids=true,
    major grid style={dashed,gray!30},
    tick style={draw=none},
    ylabel near ticks,
    axis x line*=bottom,
    axis y line*=left,
      ylabel={\footnotesize Pattern Occurrences},
  },
  observedjoin/.style={
    ybar,
    bar width=3.2pt,
    width=6.35cm,
    height=3.2cm,
    ymin=0,
    xtick=data,
    tick label style={font=\scriptsize},
    label style={font=\scriptsize},
    title style={font=\small},
    x tick label style={rotate=23, anchor=east, font=\scriptsize},
    ymajorgrids=true,
    major grid style={dashed,gray!30},
    tick style={draw=none},
    ylabel near ticks,
    axis x line*=bottom,
    axis y line*=left,
  },
  sqlancerbar/.style={draw=blue, fill=white, line width=0.45pt, bar shift=-6pt},
  eetbar/.style={draw=green!55!black, fill=white, line width=0.45pt, bar shift=-2pt},
  nomtbar/.style={draw=brown, fill=white, line width=0.45pt, bar shift=2pt},
  distrangerbar/.style={draw=red, fill=red, line width=0.45pt, bar shift=6pt},
}

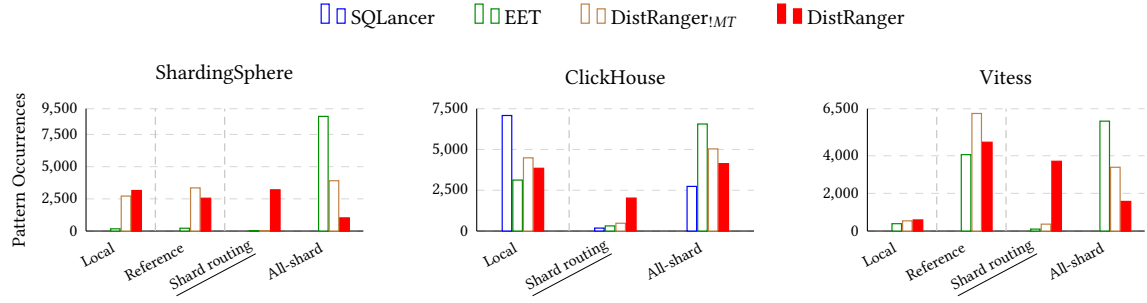
\begin{figure*}[t]
  \centering

  \begin{tikzpicture}
    \begin{groupplot}[
      group style={group size=3 by 1, horizontal sep=1.5cm},
      observedpattern,
    ]
      % -------- Citus --------
      % \nextgroupplot[
      %   title={Citus},
      %   ymax=9500,
      %   ytick={0,2500,5000,7500,9500},
      % ]
      % % x=1: Local; x=2: Reference-table access; x=3: shard-routing/single-shard access; x=4: All-shard access.
      % \addplot+[sqlancerbar] coordinates {(1,2420) (2,278) (3,313) (4,8695)};
      % \addplot+[eetbar] coordinates {(1,5251) (2,3918) (3,28) (4,4467)};
      % \addplot+[nomtbar] coordinates {(1,3921) (2,2961) (3,12) (4, 5483)};
      % \addplot+[distrangerbar] coordinates {(1,4698) (2,2153) (3,2694) (4,5697)};
      % \drawaccessseparatorsfour{9500}

      % -------- ShardingSphere --------
      \nextgroupplot[
        title={ShardingSphere},
        ymax=9500,
        ytick={0,2500,5000,7500,9500},
      ]
      \addplot+[eetbar] coordinates {(1,174) (2,219) (3,8) (4,8902)};
      \addplot+[nomtbar] coordinates {(1,2716) (2,3350) (3,25) (4,3908)};
      \addplot+[distrangerbar] coordinates {(1,3148) (2,2549) (3,3201) (4,1023)};
      \drawaccessseparatorsfour{9500}

      % -------- ClickHouse --------
      \nextgroupplot[
        title={ClickHouse},
        xmin=0.5,
        xmax=3.5,
        xtick={1,2,3},
        xticklabels={Local,\underline{Shard routing},All-shard},
        ymax=7500,
        ytick={0,2500,5000,7500},
        ylabel={},
      ]
      \addplot+[sqlancerbar] coordinates {(1,7083) (2,181) (3,2736)};
      \addplot+[eetbar] coordinates {(1,3124) (2,315) (3,6561)};
      \addplot+[nomtbar] coordinates {(1,4487) (2,478) (3,5035)};
      \addplot+[distrangerbar] coordinates {(1,3847) (2,2020) (3,4133)};
      \drawaccessseparatorone{7500}

      % -------- Vitess --------
      \nextgroupplot[
        title={Vitess},
        ymax=6500,
        ytick={0,2000,4000,6500},
        ylabel={},
      ]
      \addplot+[eetbar] coordinates {(1,390) (2,4059) (3,103) (4,5838)};
      \addplot+[nomtbar] coordinates {(1,542) (2,6246) (3,366) (4,3388)};
      \addplot+[distrangerbar] coordinates {(1,590) (2,4719) (3,3707) (4,1574)};
      \drawaccessseparatorsfour{6500}
    \end{groupplot}

    % Shared legend using PGFPlots' native ybar legend symbols.
    \coordinate (legendpos) at
      ($(group c1r1.north west)!0.5!(group c3r1.north east)+(0,12mm)$);

    \begin{axis}[
      at={(legendpos)},
      anchor=south,
      hide axis,
      xmin=0, xmax=1,
      ymin=0, ymax=1,
      width=0pt,
      height=0pt,
      scale only axis,
      ybar=1pt,
      /pgf/bar width=5pt,
      legend columns=4,
      legend style={
        at={(0.5,0.5)},
        anchor=center,
        draw=none,
        fill=none,
        font=\small,
        cells={anchor=west},
        /tikz/every even column/.append style={column sep=0.45cm},
      },
    ]
      \addlegendimage{ybar legend,draw=blue,fill=white,line width=0.45pt}
      \addlegendentry{SQLancer}

      \addlegendimage{ybar legend,draw=green!55!black,fill=white,line width=0.45pt}
      \addlegendentry{EET}

      \addlegendimage{ybar legend,draw=brown,fill=white,line width=0.45pt}
      \addlegendentry{\mytool{}$_{\mathit{!MT}}$}

      \addlegendimage{ybar legend,draw=red,fill=red,line width=0.45pt}
      \addlegendentry{\mytool{}}
    \end{axis}

  \end{tikzpicture}

  \captionsetup{skip=2pt}
  \caption{Distribution of different shard-access patterns (local, reference-table, and distributed-table access) exercised by 10K executable queries. 
  The distributed-table access patterns include shard routing and all-shard access. ClickHouse does not support reference tables.
  }
  \label{fig:shard-access-patterns2}
\end{figure*}

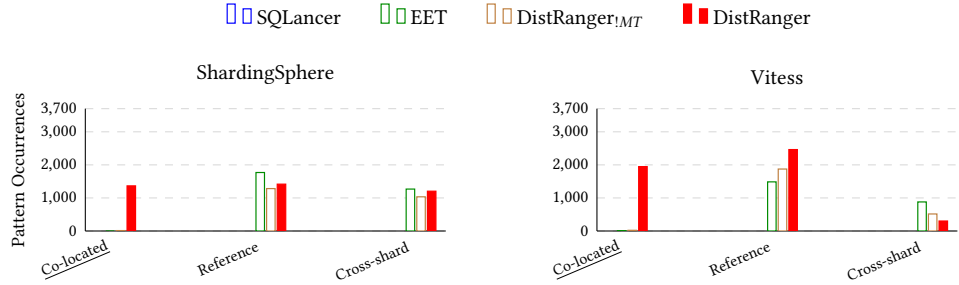
\begin{figure*}[t]
  \centering
  % {\footnotesize
  %   \patternlegendboxopen{blue} SQLancer\quad
  %   \patternlegendboxopen{green!55!black} EET\quad
  %   \patternlegendboxopen{brown} \mytool$_{\mathit{!MT}}$\quad
  %   \patternlegendbox{red} \mytool{}
  % }
  % \vspace{0.3em}

  \begin{tikzpicture}
    \begin{groupplot}[
      group style={group size=2 by 1, horizontal sep=2cm},
      observedjoin,
      symbolic x coords={Coloc.,Ref.,Distributed},
      xticklabels={\underline{Co-located},Reference ,Cross-shard},
      ymax=3700,
      ytick={0,1000,2000,3000,3700},
      ylabel={\footnotesize Pattern Occurrences},
    ]
      % -------- Citus --------
      % \nextgroupplot[title={Citus}]
      % \addplot+[sqlancerbar] coordinates {(Coloc.,1020) (Ref.,165) (Distributed,1436)};
      % \addplot+[eetbar] coordinates {(Coloc.,1169) (Ref.,1268) (Distributed,3409)};
      % \addplot+[nomtbar] coordinates {(Coloc.,622) (Ref.,377) (Distributed,3283)};
      % \addplot+[distrangerbar] coordinates {(Coloc.,2008) (Ref.,616) (Distributed,2839)};

      % -------- ShardingSphere --------
      \nextgroupplot[title={ShardingSphere}]
      \addplot+[eetbar] coordinates {(Coloc.,0) (Ref.,1769) (Distributed,1268)};
      \addplot+[nomtbar] coordinates {(Coloc.,13) (Ref.,1283) (Distributed,1034)};
      \addplot+[distrangerbar] coordinates {(Coloc.,1365) (Ref.,1417) (Distributed,1204)};

      % -------- Vitess --------
      \nextgroupplot[title={Vitess}, ylabel={}]
      \addplot+[eetbar] coordinates {(Coloc.,10) (Ref.,1486) (Distributed,879)};
      \addplot+[nomtbar] coordinates {(Coloc.,21) (Ref.,1874) (Distributed,515)};
      \addplot+[distrangerbar] coordinates {(Coloc.,1948) (Ref.,2463) (Distributed,302)};
    \end{groupplot}

    % Shared legend using PGFPlots' native ybar legend symbols.
    \coordinate (legendpos) at
      ($(group c1r1.north west)!0.5!(group c2r1.north east)+(0,12mm)$);

    \begin{axis}[
      at={(legendpos)},
      anchor=south,
      hide axis,
      xmin=0, xmax=1,
      ymin=0, ymax=1,
      width=0pt,
      height=0pt,
      scale only axis,
      ybar=1pt,
      /pgf/bar width=5pt,
      legend columns=4,
      legend style={
        at={(0.5,0.5)},
        anchor=center,
        draw=none,
        fill=none,
        font=\small,
        cells={anchor=west},
        /tikz/every even column/.append style={column sep=0.45cm},
      },
    ]
      \addlegendimage{ybar legend,draw=blue,fill=white,line width=0.45pt}
      \addlegendentry{SQLancer}

      \addlegendimage{ybar legend,draw=green!55!black,fill=white,line width=0.45pt}
      \addlegendentry{EET}

      \addlegendimage{ybar legend,draw=brown,fill=white,line width=0.45pt}
      \addlegendentry{\mytool{}$_{\mathit{!MT}}$}

      \addlegendimage{ybar legend,draw=red,fill=red,line width=0.45pt}
      \addlegendentry{\mytool{}}
    \end{axis}

  \end{tikzpicture}

  \captionsetup{skip=2pt}
  \caption{
  Distribution of different distributed-join patterns (co-located, reference, and cross-shard joins) exercised by 10K executable queries. 
ClickHouse is omitted because it does not support co-located joins.
}
  \label{fig:distributed-join-patterns2}
\end{figure*}

\subsection{Coverage of Distributed Execution Patterns}
In addition to Citus, 
we measure various shard-access and distributed-join patterns for ShardingSphere, Vitess, and ClickHouse.
For ClickHouse and Citus, we analyze \emph{query plans}, which expose accessed tables, shards (for distributed tables), and whether joins are executed in a co-located or cross-shard manner.
For ShardingSphere and Vitess, which do not expose physical plans in the same form, we use \sqlkw{PREVIEW} and \sqlkw{VEXPLAIN} to inspect rewritten or routed SQL queries and infer shard accesses and join patterns.

The experimental results are shown in Figure~\ref{fig:shard-access-patterns2} and Figure~\ref{fig:distributed-join-patterns2}.
They are consistent with the observations for Citus:
for the two target distributed optimizations, i.e., shard routing and co-located joins, significantly more queries generated by \mytool{} exercise the corresponding patterns compared with SQLancer, EET, and \mytool$_{\mathit{!MT}}$.
Additionally, due to the randomness in query generation, \mytool{} also exercises patterns beyond the two target optimizations, and each query may exercise zero or more patterns.

\onecolumn
\begin{longtable}{p{0.5cm} p{2.0cm} p{7.2cm} p{1.2cm} p{1.5cm} p{2.5cm}}
\caption{A complete list of bugs detected by \mytool{}.} \label{tab:bugs} \\
\toprule
 & DDBMS & Description & Type & Status & Features Involved \\
\midrule
\endfirsthead
\caption[]{A complete list of bugs detected by \mytool{} (Continued)} \\
\toprule
& DDBMS & Description & Type & Status & Features Involved  \\
\midrule
\endhead
\midrule
\multicolumn{6}{r}{Continued on next page} \\
\midrule
\endfoot
\bottomrule
\endlastfoot
1 & Citus & A \sqlkw{FULL OUTER JOIN} between a distributed table and a reference table produce duplicate result due to no access to \textit{set\_join\_pathlist\_hook}.  & logic & fixed & join \\
2 & Citus & An \sqlkw{UPDATE} statement with distributed table produce wrong result due to wrong subplan of \sqlkw{WHERE} filter. & logic & fixed & DML, subquery \\
3 & Citus & Subquery with reference table returns wrong result due to wrong single-shard routing logic. & logic & fixed & DML \\
4 & Citus & Adding \sqlkw{WHERE TRUE} in subquery flips result due to wrong distributed query planning. & logic & fixed & subquery \\
5 & Citus & Server crashes when re-adding a node during reset. & crash & investigating & DQL \\
6 & Citus & A \sqlkw{Delete} statement crashes due to an assertion failure in distributed query rewriting. & crash & fixed & DML, subquery \\
7 & Citus & Subquery with \sqlkw{LIMIT} returns wrong result due to wrong distributed subplan. & logic & fixed & subquery \\
8 & Citus & Segmentation fault occurs on \sqlkw{LEFT JOIN} to a distributed table and correlated subqueries. & crash & fixed & subquery, join \\
9 & ShardingSphere & Wrong result of \sqlvar{information\_schema.tables} when sharding a table. & logic & investigating & metadata \\
10 & ShardingSphere & Incorrect result window aggregate on sharded table due to wrong aggregation logic. & logic & fixed & window \\
11 & ShardingSphere & Correlated subquery in \sqlkw{HAVING} cannot reference outer query column in SQL parser. & error & fixed & subquery, group by \\
12 & ShardingSphere & Exception when \sqlkw{SELECT} version() due to wrong function verify and routing. & exception & fixed & / \\
13 & ShardingSphere & Timestamp output adds trailing ".0" in SQL format. & logic & fixed & / \\
14 & ShardingSphere & Correlated subquery triggers timeout due to wrong distributed query rewriting.  & timeout & confirmed & subquery, join \\
15 & ShardingSphere & Correlated \sqlkw{IN}-subquery fails in a non-join subquery due to wrong distributed planing logic. & error & fixed & subquery, join \\
16 & ShardingSphere & Co-located join inside scalar subquery fails due to wrong distributed query rewriting. & error & fixed & join, subquery \\
17 & ShardingSphere & Result of \sqlkw{SELECT DISTINCT} duplicates due to wrong distributed query rewriting. & logic & fixed & distinct \\
18 & ClickHouse & Nested \sqlkw{EXISTS} with deep correlated reference fails. & exception & confirmed & subquery \\
19 & ClickHouse & Result of \sqlkw{GLOBAL RIGHT OUTER JOIN} on distributed tables duplicates. & logic & investigating & join \\
20 & Vitess & ApplyVSchema accepts empty auto\_increment object and later \sqlkw{INSERT} fails. & error & investigating & shard \\
21 & Vitess & Cannot insert into pinned table, got no primary vindex due to wrong distributed query routing. & error & confirmed & shard \\
22 & Vitess & Window function without single-shard restriction returns wrong result with uncorrelated scalar subquery. & logic & investigating & subquery, window \\
23 & Vitess & VTGate panic when \sqlkw{WHERE} contains \sqlkw{LTRIM} on sharded tables. & crash & investigating & subquery, join \\
24 & Vitess & VTGate planning appears to hang on queries with many uncorrelated scalar subqueries. & timeout & investigating & subquery, join \\
25 & Vitess & subquery is incorrectly dropped on sharded keyspace. & logic & investigating & subquery \\
26 & Vitess & Unknown column in \sqlkw{WHERE} clause of scalar subquery. & error & investigating & subquery,  \\

 & &  &  &  & aggregation \\

27 & Vitess & Planner drops constant-true \sqlkw{OR} branch and returns wrong result on sharded keyspace. & logic & fixed & subquery \\
28 & Vitess & VTGate panic in planner when comparing two scalar subqueries. & crash & confirmed & subquery \\
29 & Vitess & Reference table \sqlkw{SELECT} incorrectly rewritten to \sqlkw{WHERE} 1!=1 with nested \sqlkw{EXISTS}. & logic & confirmed & subquery \\
30 & Vitess & Reference table \sqlkw{OUTER JOIN} on preserved side duplicates rows across shards. & logic & investigating & join \\
31 & Vitess & Co-located join rewrite breaks \sqlkw{LEFT OUTER JOIN} semantics due to wrong distributed query rewriting. & logic & confirmed & join \\
\end{longtable}
\twocolumn

\end{document}